\documentclass[useAMS,usenatbib]{aa/aa}
\usepackage{natbib}
\bibpunct{(}{)}{;}{a}{}{,} 
\usepackage{graphicx}
\usepackage{txfonts}
\usepackage{hyperref}
\usepackage{epstopdf}
\usepackage{amsmath, amssymb}
\def\apjl{ApJ}%
\def\apjs{ApJS}%
\def\ao{Appl.~Opt.}%
\def\aap{A\&A}%
\newcommand {\reac}[6] {$\rm\,{}^{#2}\kern-0.8pt{#1}\,({#3}\,,{#4}) \,{}^{#6}\kern-0.8pt{#5}\,$}

\newcommand{\feh}{\mbox{\rm [{\rm Fe}/{\rm H}]}}
\newcommand{\mh}{\mbox{\rm [{\rm M}/{\rm H}]}}
\newcommand{\Msun}{\mbox{$M_{\odot}$}}

\newcommand{\logL}{\mbox{$\log L/L_{\odot}$}}

\newcommand{\logl}{\mbox{$\log L/L_{\odot}$}}
\newcommand{\logg}{\mbox{$\log g$}}

\newcommand{\beq}{\begin{equation}}
\newcommand{\eeq}{\end{equation}}
\newcommand{\beqa}{\begin{eqnarray}}
\newcommand{\eeqa}{\end{eqnarray}}

\newcommand{\BC}{{\rm BC}}
\newcommand{\Av}{A_{V}}
\newcommand{\teff}{T_{\rm eff}}
\newcommand{\logteff}{{\rm log}\,T_{\rm eff}}

\begin{document}

\title{YBC, a stellar bolometric corrections database with variable extinction coefficients: an application to {\it PARSEC} isochrones}
\titlerunning{YBC bolometric corrections database}

\author{Yang Chen\inst{1}\fnmsep\thanks{\email{yang.chen@unipd.it}}
  \and
  L\'eo Girardi\inst{2}
  \and
  Xiaoting Fu\inst{3}
  \and
  Alessandro Bressan\inst{4}
  \and
  Bernhard Aringer\inst{1}
  \and
  Piero Dal Tio\inst{1,2}
  \and
  Giada~Pastorelli\inst{1}
  \and
  Paola Marigo\inst{1}
  \and
  Guglielmo Costa\inst{4}
  \and
  Xing Zhang\inst{5}
}

\authorrunning{Chen et al.}

\institute{
  Dipartimento di Fisica e Astronomia Galileo Galilei, Universit\`a di Padova, Vicolo dell'Osservatorio 3, I-35122 Padova, Italy
  \and
  Osservatorio Astronomico di Padova -- INAF, Vicolo dell'Osservatorio 5, I-35122 Padova, Italy
  \and
  The Kavli Institute for Astronomy and Astrophysics at Peking University, Beijing, China
  \and
  SISSA, via Bonomea 365, I-34136 Trieste, Italy
  \and
  Department of Astronomy, University of Science and Technology of China, Chinese Academy of Sciences, Hefei, Anhui 230026, China
}
\date{Received - ; accepted -}

\hypersetup{
    linkcolor=blue,
    citecolor=blue,
    filecolor=magenta,      
    urlcolor=cyan
}

\date{Received .... / Accepted .....}

\abstract{
  We present the \texttt{YBC} database of stellar bolometric corrections
  (BCs), available at \url{http://stev.oapd.inaf.it/YBC}. We
  homogenise widely-used theoretical stellar spectral libraries and
  provide BCs for many popular photometric systems, including the Gaia
  filters. The database can be easily extended to additional
  photometric systems and stellar spectral libraries. The web
  interface allows users to transform their catalogue of theoretical
  stellar parameters into magnitudes and colours of selected filter
  sets. The BC tables can be downloaded or also be implemented into
  large simulation projects using the interpolation code provided with
  the database. We compute extinction coefficients on a star-by-star
  basis, hence taking into account the effects of spectral type and
  non-linearity dependency on the total extinction. We illustrate the 
  use of these BCs in \texttt{PARSEC} isochrones. We show that using
  spectral-type dependent extinction coefficients is quite necessary
  for Gaia filters whenever $\Av\gtrsim 0.5$\,mag. BC tables for
  rotating stars and tables of limb-darkening coefficients are also
  provided.
  }

\maketitle

\section{Introduction}
\label{intro}

Bolometric corrections (BCs) are usually applied to the absolute
magnitude of a star to obtain its bolometric absolute magnitude or
luminosity, or conversely, to predict the magnitudes of a model star in
a given set of filters.
Most modern versions of BCs are usually based on theoretical stellar
spectral libraries, e.g.
\citet{Girardi2002,Casagrande2014,Casagrande2018a}. In some cases
empirical stellar spectral libraries are also used for this purpose,
however, they require additional calibration from the theoretical
atmosphere models and are always limited by the coverage of stellar
spectral type and wavelength range
\citep{MILES2006,IRTF2009,XSL2014}. In practice, BCs are often
provided along with, or can be easily computed from the spectral
libraries obtained from the modelling of stellar atmospheres. In the
present day, the widely used theoretical stellar spectral libraries
include \texttt{ATLAS9} \citep{ATLAS9} and PHOENIX \citep{PHOENIX} for
generic types of stars, MARCS \citep{MARCS} for cool or intermediate
temperature stars, \texttt{COMARCS} for K/M/S/C stars
\citep{COMARCS_C,COMARCS_M,Aringer2019}, and \texttt{PoWR} for
Wolf-Rayet (WR) stars \citep[][and refs. therein]{PoWR2002}. These
libraries, with different solar abundance, offer large grids of models
covering different stellar parameters (metallicity, $\teff$ and
$\logg$).  However, currently there is a lack of a single,
homogeneous, public database to synthesise these libraries and provide
BC tables for a large set of photometric systems.

One of the obvious advantages of having a homogeneous database for BCs
would be to facilitate the comparison between stellar models and the
observations. Indeed, besides the differences in the theoretical
quantities such as $\logl$ and $\teff$ of various stellar models,
additional deviations are often introduced from different authors
using different BC tables.  The possibility of testing the difference
from those originally applied will become more critical
with forthcoming facilities providing photometry of much higher
precision: considering, for instance, the $\sim10^{-4}$~mag errors
expected for Gaia sources with $G<12$~mag and $>100$ CCD transits, 
for a nominal mission with perfect calibrations
\citep[see figure 9 in][]{evans18}, and the systematic errors smaller than 0.005~mag
planned for LSST single visits \citep{ivezic19}. Another aspect to
consider is the different ways of handling the interstellar extinction
in the models, which is especially crucial when using either UV
filters (as in GALEX) or very broad passbands (such as the Gaia, TESS,
and HST/WFC3/UVIS extremely wide filters). In these cases, fixed
extinction coefficients are no longer valid and the effect of stellar
spectral types becomes critical \citep{Girardi2008}.

In this work, we build a database where we assemble existing popular
stellar spectral libraries to compute BC tables homogeneously for a
wide variety of photometric systems. The web interface of this
database provides a convenient way for the users to transform their
theoretical stellar catalogues into magnitudes and colours, and hence to
compare them with observations. It has the flexibility to choose
different stellar spectral libraries, thus allowing their differences
to be easily investigated. The extinction coefficients have been
computed on a star-by-star basis, therefore the variation with
spectral type has been taken into account. The non-linearity as a
function of $\Av$ has been included, which is important for highly
attenuated targets.

In this paper, we introduce the definitions of the BCs in section
\ref{sec:bc}, the available stellar spectral libraries in section
\ref{sec:spectra} and our C/Python code package in section
\ref{sec:package}. 
We compare some of our results with the literature in section \ref{sec:bccomparison}.
In section \ref{sec:extinction}, we discuss the
spectral type dependent extinction coefficients. Section
\ref{sec:discussion} summarises the main results.

\section{Bolometric corrections}
\label{sec:bc}
In this section, we recall some basic relations concerning the
bolometric correction which are necessary for the discussion. We refer
the readers to \citet{Kurucz1979}, \citet{Bessell1998}, and 
\citet{Girardi2002} for a more exhaustive discussion.

First, we recall the definition of magnitude. Assuming we (at earth)
receive a radiation flux $f_\lambda$ (or $f_\nu$) from a source, the
magnitude in a certain filter band $i$ with transmission curve
$S_{\lambda,i}$ is
\begin{equation}
  m_i=-2.5{\rm log}\left[\frac{\int_{\lambda_1}^{\lambda_2} \lambda f_{\lambda} S_{\lambda, i} d\lambda}
    {\int_{\lambda_1}^{\lambda_2} \lambda f^0_\lambda S_{\lambda, i} d\lambda}\right] + m_{i,0}.
\end{equation}
In this equation, $f^0_\lambda$ is the flux of the reference spectrum
and $m_{i,0}$ is the corresponding reference magnitude.  These two
quantities depend on the photometric system and will be discussed 
later. 
$\lambda_1$ and $\lambda_2$ denote the lower and upper wavelength limits of the filter transmission curve, respectively.
The above equation is valid for present-day photon-counting
devices (CCDs or IR arrays). For more traditional energy-integrating
systems the above equation should be changed to
\begin{equation}
  m_i=-2.5{\rm log}\left[\frac{\int_{\lambda_1}^{\lambda_2} f_{\lambda} S_{\lambda, i} d\lambda}
    {\int_{\lambda_1}^{\lambda_2} f^0_\lambda S_{\lambda, i} d\lambda}\right] + m_{i,0}.
\end{equation}

Depending on the reference spectra $f^0_\lambda$ (or $f^0_\nu$),
commonly used magnitude systems are:
\begin{itemize}
\item Vega magnitude systems. The spectrum of Vega ($\alpha$ Lyr) is
  used as the reference spectrum. The reference magnitudes are set so
  that Vega has a magnitude equal to, or slightly different from
  zero. By default, we use the latest Vega spectrum\footnote{Currently
    it is \url{alpha_lyr_stis_008.fits} from
    \url{ftp://ftp.stsci.edu/cdbs/current_calspec/}.} from the CALSPEC
  database \citep{Bohlin2014}. 

\item AB magnitude systems \citep{Oke1974}. The reference spectrum has
  a constant value of $f^0_\nu=10^\frac{48.60}{-2.5}$ $\rm
  erg\ s^{-1}\ cm^{-2}\ Hz^{-1}$. The reference magnitudes thus are
  set to $-$48.60 mag.

\item Space Telescope (ST) magnitude systems \citep{Laidler2005}. The
  reference spectrum has a constant value of
  $f^0_\lambda=10^\frac{21.1}{-2.5}$ $\rm
  erg\ s^{-1}\ cm^{-2}\ \text{\AA}^{-1}$. The reference magnitudes
  thus are set to $-$21.10 mag.  

\item Gunn systems \citep{TG1976}. In these systems, F-type subdwarfs,
  in particular BD +17 4708, are taken as the reference stars instead
  of Vega.
 
\end{itemize}
Among these, the Vega and AB magnitude systems are the most widely adopted ones.
At \url{http://stev.oapd.inaf.it/cmd/photsys.html}, the reader can 
check the information about the photometric systems supported.

Usually, synthetic  spectral libraries provide the stellar flux at the stellar
radius $R$. This flux $F_\lambda$ is related to the effective
temperature $T_{\rm eff}$ of the star by
\begin{equation}
  F_{\rm bol} \equiv \int_0^\infty F_\lambda d\lambda=\sigma T^4_{\rm eff},
  \label{eq_Fbol}
\end{equation}
where $\sigma$ is the Stefan-Boltzmann constant. By placing the star
at 10 pc from the earth, the flux we receive is
\begin{equation*}
  f_{\lambda, \rm 10 pc}=\left(\frac{R}{\rm 10 pc}\right)^2 F_\lambda 10^{-0.4A_\lambda},
\end{equation*}
where $A_\lambda$ is the assumed extinction between the star and the observer. Therefore, the
absolute magnitude $M_i$ for a photon-counting photometric system is
\begin{equation}
  \begin{split}
    M_i&=-2.5{\rm log}\left[\frac{\int_{\lambda_1}^{\lambda_2} \lambda f_{\lambda, \rm 10 pc} S_{\lambda, i} d\lambda}
      {\int_{\lambda_1}^{\lambda_2} \lambda f^0_\lambda S_{\lambda, i} d\lambda}\right] + m_{i,0}\\
    &=-2.5{\rm log}\left[ \left(\frac{R}{\rm 10 pc}\right)^2
      \frac{\int_{\lambda_1}^{\lambda_2} \lambda F_\lambda 10^{-0.4A_\lambda} S_{\lambda, i} d\lambda}
           {\int_{\lambda_1}^{\lambda_2} \lambda f^0_\lambda S_{\lambda, i} d\lambda}\right] + m_{i,0}.
  \end{split}
  \label{eq_Mag}
\end{equation}
The definition of bolometric magnitude $M_{\rm bol}$ is
\begin{equation}
  \begin{split}
    M_{\rm bol} &= M_{\rm bol,\odot} - 2.5{\rm log} (L/L_{\rm \odot})\\
    &= M_{\rm bol,\odot} - 2.5{\rm log} (4\pi R^2F_{\rm bol}/L_{\rm \odot}).
  \end{split}
  \label{eq_Mbol}
\end{equation}
According to the IAU 2015 resolution \citep{IAU2015}, the absolute
bolometric magnitude for the nominal solar luminosity ($3.828\times10^{26}$ W)
is $M_{\rm bol,\odot}= 4.74$ mag.

Given an absolute magnitude $M_i$ in a given filter band $i$ for a
star of absolute bolometric magnitude $M_{\rm bol}$, the bolometric
correction $\BC_i$ is:
\begin{equation}
  \BC_i=M_{\rm bol}- M_i.
  \label{eq_BC}
\end{equation}
By combing equations (\ref{eq_Mag}), (\ref{eq_Mbol}) and (\ref{eq_BC}), we have 
\begin{equation} 
  \begin{split}
    \BC_i=&M_{\rm bol,\odot}-2.5{\rm log} \left( \frac{4\pi \sigma ({\rm 10 pc})^2}{L_{\rm \odot}} \right) - 2.5{\rm log} (T^4_{\rm eff}) \\
    &+2.5{\rm log}\left[ \frac{\int_{\lambda_1}^{\lambda_2} \lambda F_\lambda 10^{-0.4A_\lambda} S_{\lambda, i} d\lambda}
      {\int_{\lambda_1}^{\lambda_2} \lambda f^0_\lambda S_{\lambda, i} d\lambda}\right] - m_{i,0}. 
  \end{split}
  \label{eq_BC_photon}
\end{equation}
The advantage of using the above equation to compute BCs is that
the stellar radius $R$ disappears.
In some cases, the stellar spectral library is
computed in plane-parallel geometry and there is no definition of 
a geometrical stellar radius (but the optical depth). Therefore, once
$F_\lambda$ (related to $T_{\rm eff}$ by equation (\ref{eq_Fbol})) for
a star with given $T_{\rm eff}$, $\logg$ and metallicity ${\rm [M/H]}$ is
provided, the corresponding $\BC_i$ can be derived.

The above definition of the BC is valid for present-day photon-counting
devices (CCDs or IR arrays), while for energy-integrating systems the above
equation should be changed to
\begin{equation}
  \begin{split}
    \BC_i=&M_{\rm bol,\odot}-2.5{\rm log}\left( \frac{4\pi \sigma ({\rm 10 pc})^2}{L_{\rm \odot}} \right) - 2.5{\rm log} (T^4_{\rm eff})\\
    &+2.5{\rm log}\left[ \frac{\int_{\lambda_1}^{\lambda_2} F_\lambda 10^{-0.4A_\lambda} S_{\lambda, i} d\lambda}
      {\int_{\lambda_1}^{\lambda_2} f^0_\lambda S_{\lambda, i} d\lambda}\right] - m_{i,0}. 
  \end{split}
  \label{eq_BC_energy}
\end{equation}
For AB magnitude systems with photon-counting devices,
we can either convert $f^0_\nu$ to $f^0_\lambda$ and use equation
(\ref{eq_BC_photon}), or use the following equation instead:
\begin{equation}
  \begin{split}
    \BC_i=&M_{\rm bol,\odot}-2.5{\rm log}\left( \frac{4\pi \sigma ({\rm 10 pc})^2}{L_{\rm \odot}} \right) - 2.5{\rm log} (T^4_{\rm eff})\\
    &+2.5{\rm log}\left[ \frac{\int_{\nu_1}^{\nu_2} F_\nu 10^{-0.4A_\nu} S_{\nu, i} d\nu/\nu}
      {\int_{\nu_1}^{\nu_2} f^0_\nu S_{\nu, i} d\nu/\nu}\right] - m_{i,0}. 
  \end{split}
  \label{eq_BC_photon_AB}
\end{equation}
Similarly, the equation for AB magnitude system with energy-integrating devices is:
\begin{equation}
  \begin{split}
    \BC_i=&M_{\rm bol,\odot}-2.5{\rm log}\left( \frac{4\pi \sigma ({\rm 10 pc})^2}{L_{\rm \odot}} \right) - 2.5{\rm log} (T^4_{\rm eff})\\
    &+2.5{\rm log}\left[ \frac{\int_{\nu_1}^{\nu_2} F_\nu 10^{-0.4A_\nu} S_{\nu, i} d\nu}
      {\int_{\nu_1}^{\nu_2} f^0_\nu S_{\nu, i} d\nu}\right] - m_{i,0}. 
  \end{split}
  \label{eq_BC_energy_AB}
\end{equation}

\section{Stellar spectral libraries}
\label{sec:spectra}

In this section, we briefly describe the stellar spectral libraries
currently supported in our database.  We will continuously expand it 
with data from external sources or provided by our group.

\subsection{\texttt{ATLAS9} models}
One of the most widely used stellar spectral libraries is the plane
parallel models computed by
\citet{ATLAS9}\footnote{\url{http://wwwuser.oats.inaf.it/castelli/grids.html}.}
with the \texttt{ATLAS9} code \citep{ATLAS9&12}.  These models are
based on the solar abundances by \cite{GS98} and make use of an
improved set of molecular absorption lines including TiO and $\rm
H_2O$, as well as absorption lines from quasi-molecular H-H and
H-H$^+$.  The model grids are computed for $\teff$ from 3500\,K to
50000\,K, $\logg$ ($g$ in cgs unit) from 0.0 dex to 5.0 dex and
[M/H]=+0.5, +0.2, 0.0, $-$0.5, $-$1.0, $-$1.5, $-$2.0, $-$2.5, $-$3.5,
$-$4 and $-$5.5.
    
\subsection{\texttt{PHOENIX} models}
The \texttt{PHOENIX} database \citep{Allard1995,Allard1997} is another
widely used stellar spectra library, especially for cool stars. The
atmospheres of cool stars are dominated by molecular formation and by
dust condensation at very low temperature. Both of these two phenomena
can affect the spectral shape significantly. 
A suitable set of 1D, static spherical atmosphere
spectral models accounting for the above effects has been provided in
\citet{PHOENIX}. Among the different model suites of this database, we
use the
BT-Settl\footnote{\url{https://phoenix.ens-lyon.fr/Grids/BT-Settl/AGSS2009/}.}
models for the coverage completeness in stellar parameters and for the
wide usage in the literature. The BT-Settl models use the 
\citet{AGSS2009} solar abundances. They are provided
for $2600\,$K $\leq \teff < 50000\,K$, $0.5 < \logg < 6$, and
metallicities $-4.0\lesssim {\rm [M/H]} \lesssim +0.5$.

\subsection{\texttt{WM-basic} models}
For temperatures typical of O and B stars (19000\,K$<\teff<$60000\,K)
we have computed a library of models using the public code
\texttt{WM-basic} \citep{WMbasic}, as described in
\citet{Chen2015}. This allows us to consider both the effects
of extended winds and those of non-LTE, because both effects may
significantly affect the emergent spectra of hot stars. 
These models
use the solar abundances of \texttt{PARSEC} \citep{PARSEC}, which compiled the results 
from \cite{GS98}, \citet{Caffau2011} and references therein. The model grids are computed for 
metallicities of $Z=0.0001$, 0.0005, 0.001, 0.004, 0.008 and
0.02.  $\logteff$ covers the interval from 4.3 to 5 with a step of
0.025 dex, while $\logg$ goes from 2.5 to 6.0 with a step of 0.5
dex. For each $\logteff$ and $\logg$, models with three values of mass
loss rate ($\dot{M}=10^{-7}$, $10^{-6}$, and $10^{-5}\,
M_{\odot}\,\mathrm{yr}^{-1}$) are computed if convergence is reached
for them, as detailed in \citet{Chen2015}.

\subsection{\texttt{PoWR} models}
Wolf-Rayet (WR) stars typically have wind densities one order of magnitude larger
than those of massive O-type stars. Spectroscopically they are
dominated by the presence of strong broad emission lines of helium,
nitrogen, carbon and oxygen. They are subdivided into different
sub-types, one with strong lines of helium and nitrogen (WN stars),
another one with strong lines of helium and carbon (WC stars) and the
third one with strong oxygen lines (WO stars).  The
\texttt{PoWR}\footnote{\url{http://www.astro.physik.uni-potsdam.de/~wrh/PoWR/powrgrid1.php}.}
group has provided models for WR stars \citep{PoWR2002,
  PoWR2004, PoWR2012, PoWR2015}. These models 
  adopt the solar abundances from \citet{GS98}, and are computed for Milky
Way, Large Magellanic Cloud (LMC), Small Magellanic Cloud (SMC), and sub-SMC metallicities. For each metallicity,
late-type WN stars of different hydrogen fraction, early-type WN
stars, and WC (except for SMC and sub-SMC at the moment) stars are
computed. For a given spectral type and metallicity, the models are
computed as a function of $\logteff$ and the transformed radius $\log
R_\mathrm{t}$ \citep{Schmutz1989} (which is a more important quantity
than $\logg$ in these models dominated by stellar winds). The coverage
in $\logteff$ ($\sim4.5-5.2$) and $\log R_\mathrm{t}$ ($\sim0.-1.6$)
changes with metallicity and spectral type. The transformed radius is
defined as:
\begin{equation}
  R_\mathrm{t}=R_* \left(\frac{v_\infty}{\rm 2500~ km\,\mathrm{s}^{-1}} \middle/
  \frac{\dot{M}}{10^{-4}~{\rm M_\odot\,\mathrm{yr}^{-1}}}\right)^{2/3},
\end{equation}
with $R_*$, $v_\infty$ and $\dot{M}$ being the hydro-static stellar
radius, terminal velocity and mass loss rate, respectively.  Notice
that in the original definition of $R_\mathrm{t}$ of \texttt{PoWR}, a
multiplication factor $D^{-1/3}$ is applied to the above equation,
where $D$ is the clumping factor.  For different types of WR models of
\texttt{PoWR}, different clumping factors have been used. To
facilitate the interpolation, we have divided the $R_\mathrm{t}$
values from \texttt{PoWR} by this factor, therefore it disappears in the above equation.

\subsection{\texttt{COMARCS} models}
The local thermodynamic equilibrium (LTE) \& chemical equilibrium without dust, spherical symmetric,
hydrostatic
\texttt{COMARCS}\footnote{\url{http://stev.oapd.inaf.it/atm}.}$^,$\footnote{\url{http://stev.oapd.inaf.it/synphot/Cstars}.}
database provides models for C, S, K and M type stars for modelling
TP-AGB stars and other red giants
\citep{COMARCS_C,COMARCS_M,Aringer2019}. The solar chemical composition is based on \citet{Caffau2009a,Caffau2009b}.
The carbon star grids cover
a metallicity Z range from 0.001 to 0.016, $\teff$ from 2500\,K to
4000\,K and $\logg$ from $-$1 to 2, while the K/M star grids cover a
metallicity range from 0.00016 to 0.14, $\teff$ from 2600\,K to
4800\,K and $\logg$ from $-$1 to 5.

\subsection{DA white dwarf spectral libraries}
Gaia DR2 parallaxes \citep{Luri2018} allowed to place a quite significant
number of white dwarfs into absolute magnitude versus colour diagrams,
hence ensuring a wide interest in including these objects in
isochrone-fitting methods.  Therefore we include the
\citet{Koester2010} and \citet{Tremblay2009} DA white dwarf libraries (downloaded
from \url{http://svo2.cab.inta-csic.es/theory/newov/}) in our database.
In these plane-parallel models LTE and hydro-static equilibrium are assumed.  The
library covers the $\teff$ range from 5000\,K to 80000\,K, $\logg$ ranges
from 6.5 to 9.5.

\subsection{\texttt{ATLAS12} models with $\alpha$-enhancement}
For studying the photometric properties of alpha-enhanced
\texttt{PARSEC} stellar models, we have computed atmosphere models
with the latest \texttt{ATLAS12}
code\footnote{\url{https://www.oact.inaf.it/castelli/castelli/sources/atlas12.html}.}
with an updated line lists. These models have been used in
\citet{Fu2018}, and have been shown to improve the fitting to the
Galactic Globular Cluster 47Tuc observations with HST ACS/WFC3.  For
the moment, we only computed these spectra for two metallicities
corresponding to two 47 Tuc stellar populations \citep{Fu2018},
however, work is in progress to extend them to other metallicities
together with the \texttt{PARSEC} alpha-enhanced stellar tracks. These
models cover $\teff$ from 4000\,K to 21000\,K, and $\logg$ from 0.5 to
5. Below 10000\,K, the step in $\teff$ is 100\,K, while above the step
is 200\,K. The step in $\logg$ is 0.5 dex.

\subsection{Spectra of rotating stars based on Kurucz models}
In \citet{Girardi2019}, we have computed a spectral library for
rotating stars with a new approach. This approach takes into account
the effect of the stellar surface effective temperature variation due
to rotation -- which results in the photometric differences when the
star is observed from different angles -- as well as the
limb-darkening effect. This library is based on the spectral intensity
library from Kurucz \url{http://www.kurucz.havard.edu}. The Kurucz
models cover $\teff$ from 3500\,K to 50000\,K, $\logg$ from 0 to 5,
and $\mh$ from $-$5 to +1. The wavelength coverage is the same as that
of \texttt{ATLAS9} models, from 90.9~\AA~ to 16 $\mu$m.

\subsection{Tables of limb-darkening coefficients}
The limb-darkening effect is important to analyse light curves of
eclipsing binaries and microlensing events, and for probing exoplanet
atmospheres and spatially resolved stellar surfaces. Accurate
empirical determination of limb-darkening coefficients through
eclipsing binaries observations is not possible yet, mainly because of
parameter degeneracies. Therefore, stellar atmosphere models are
essential in these studies.  Based on the spectral intensity library
from the Kurucz website (\url{http://www.kurucz.havard.edu}), we
computed the specific intensity $I(\mu)$ as a function of
$\mu=\cos\theta$, where $\theta$ is the angle between the line of
sight and the surface normal, for each of the filter systems we have.
Specific intensities have been computed for both photon-counting and
energy-integrating detectors respectively as
\begin{equation}
  I_i(\mu) = \frac{\int_\lambda I_\lambda(\mu)\,S_{\lambda,i}\,\lambda\, \mathrm{d}\lambda}{\int_\lambda S_{\lambda,i}\,\lambda\, \mathrm{d}\lambda}
\end{equation}
and
\begin{equation}
  I_i(\mu) = \frac{\int_\lambda I_\lambda(\mu)\,S_{\lambda,i}\, \mathrm{d}\lambda}{\int_\lambda S_{\lambda,i}\, \mathrm{d}\lambda}.
\end{equation}
We further computed the limb-darkening coefficients ($a_1$, $a_2$,
$a_3$ and $a_4$) with the fitting relation proposed by
\citet{Claret2000}:
\begin{equation}
  \frac{I(\mu)}{I(1)} = 1 - \sum^4_{n=1}a_n\,\left(1-\mu^{n/2}\right).
\end{equation}
These coefficients are provided mainly for a question of completeness.
Indeed, they are computed homogeneously for all filter sets 
included in our database.

\begin{figure*}
  \includegraphics[width=0.5\textwidth]{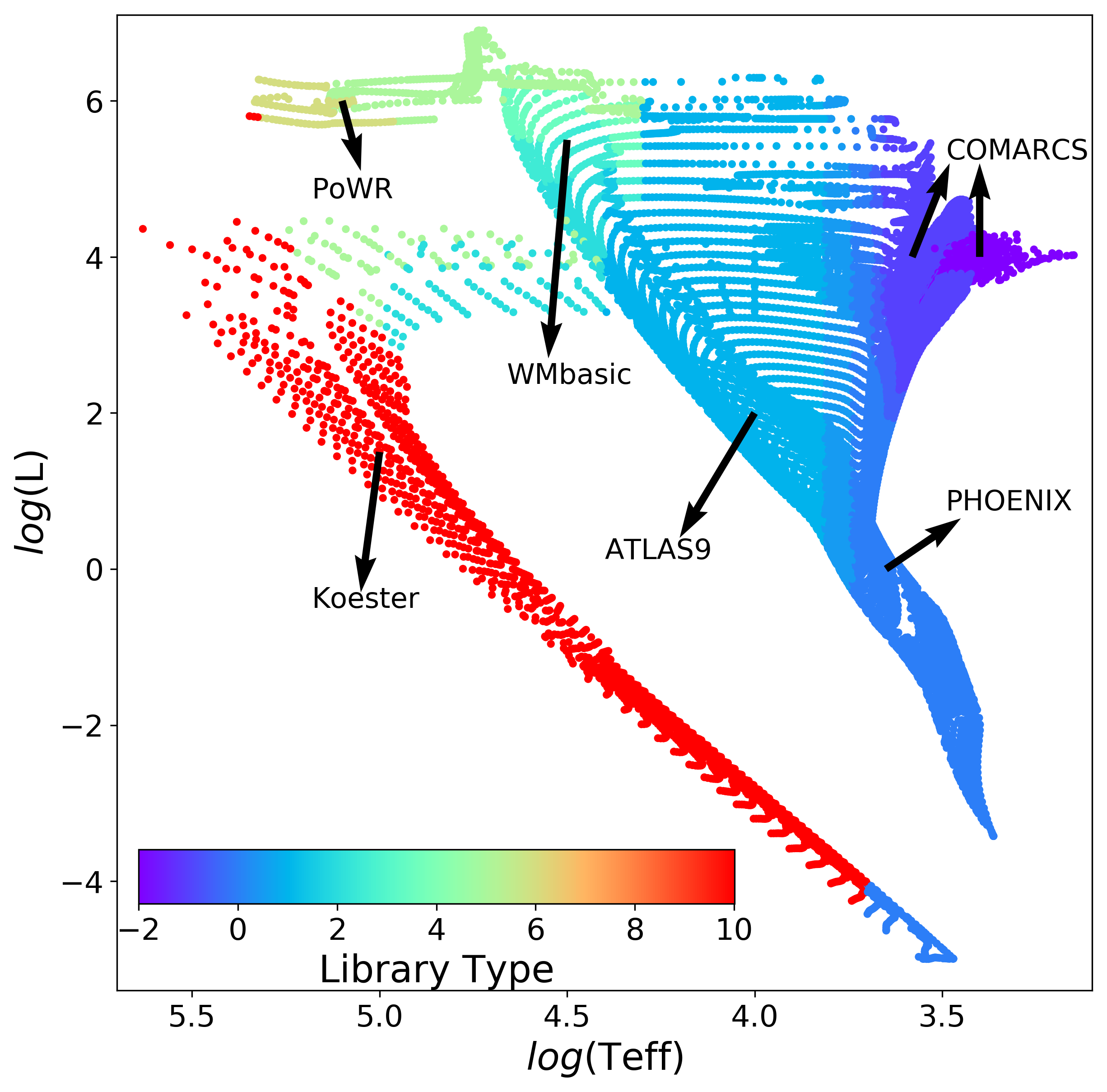}
  \includegraphics[width=0.5\textwidth]{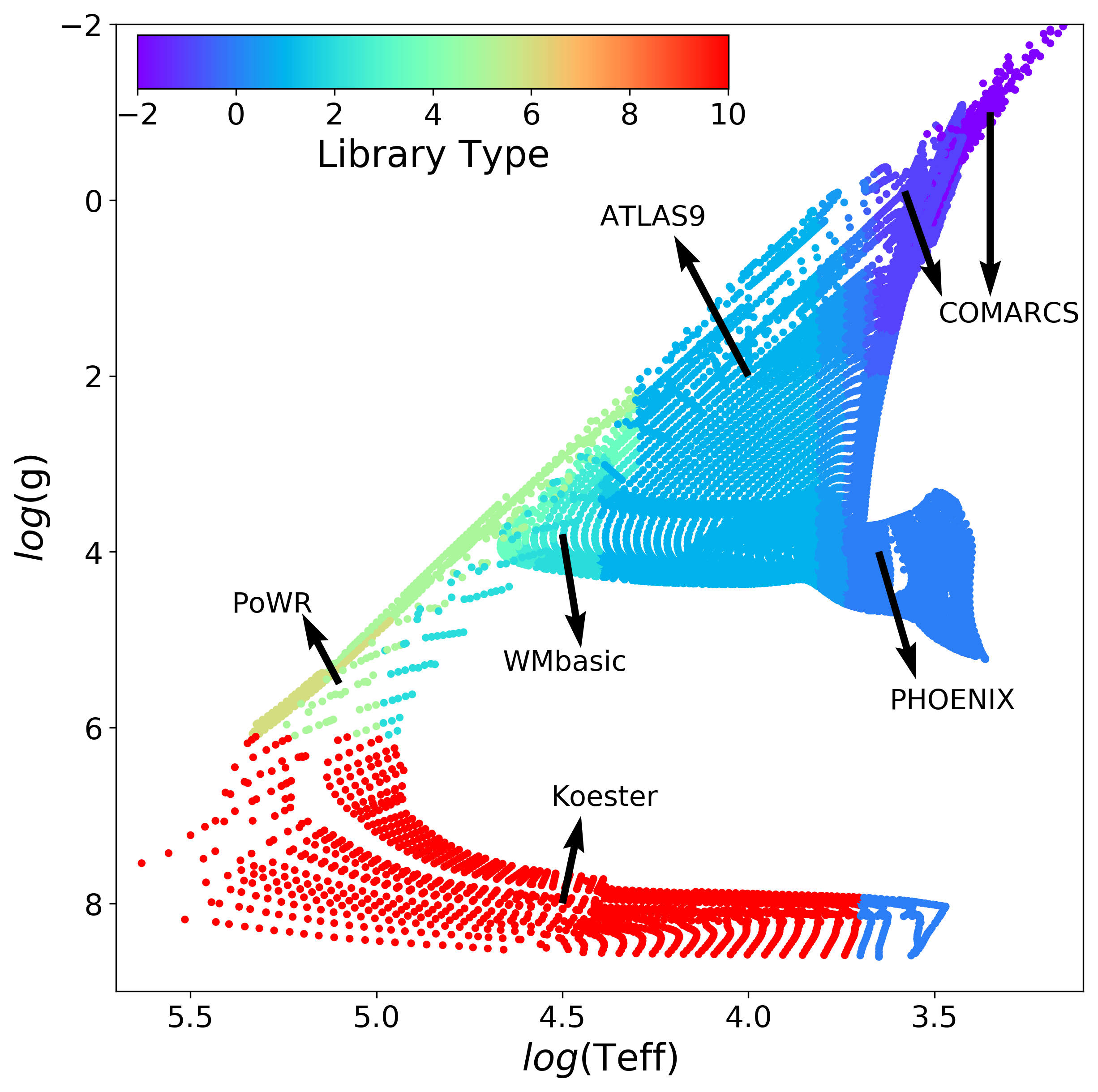}
  \includegraphics[width=0.5\textwidth]{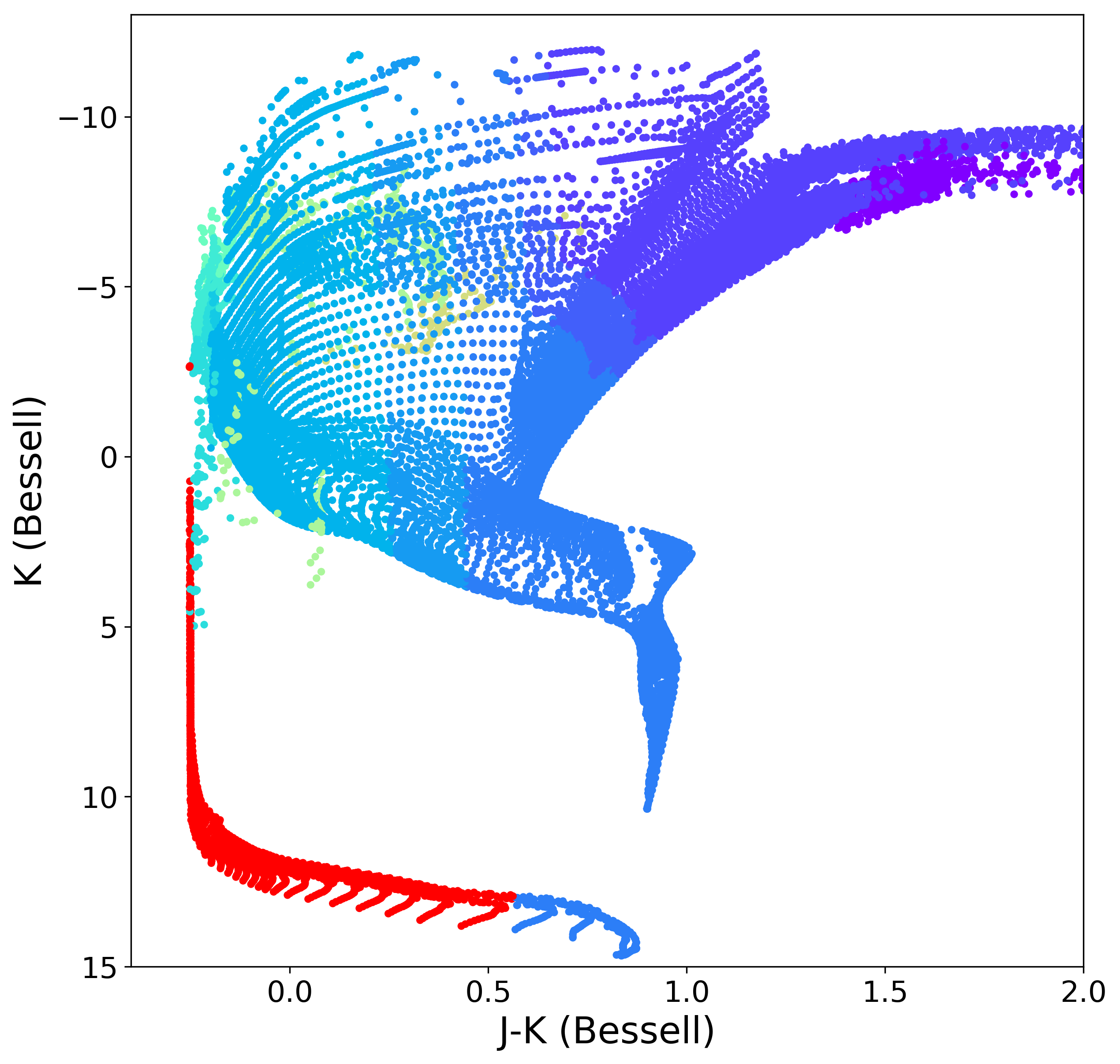}
  \includegraphics[width=0.5\textwidth]{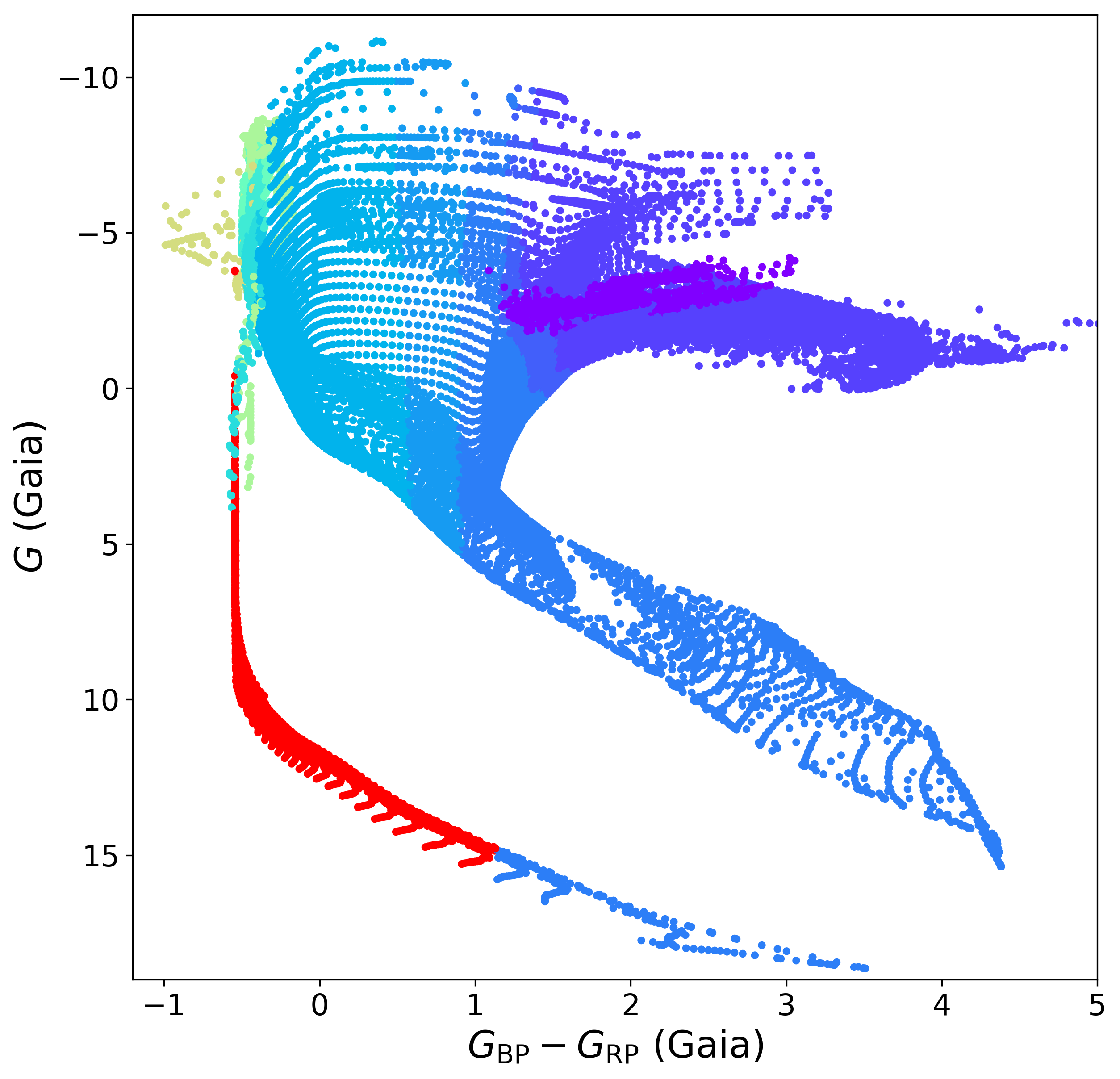}
  \caption{\texttt{PARSEC} v1.2S isochrones of [M/H]=0.0 and ages from
    log(age/yr)=6.6 to 10.1, complete from pre-main sequence to white
    dwarf stages, or up to the carbon-ignition in the case of massive stars.
    Upper left: The Hertzsprung-Russell diagram $\logL$ .vs. $\logteff$ diagram; Upper right: the Kiel diagram $\logg$ .vs. $\logteff$;
    Lower left: CMD in Bessell $JK$; 
    Lower right: CMD in Gaia DR2 filters.
    The ``library type'' we
    assign to different libraries are -2 for \texttt{COMARCS} C star
    library, -1 for \texttt{COMARCS} M/S star library, 0 for
    \texttt{PHOENIX} library, 1 for \texttt{ATLAS9} or
    \texttt{ATLAS12} library, 2 for \texttt{WM-basic} library with
    $\dot{M}=10^{-7}\,M_{\odot}\,\mathrm{yr}^{-1}$, 3 for
    \texttt{WM-basic} library with
    $\dot{M}=10^{-6}\,M_{\odot}\,\mathrm{yr}^{-1}$, 4 for
    \texttt{WM-basic} library with
    $\dot{M}=10^{-5}\,M_{\odot}\,\mathrm{yr}^{-1}$, 5 for
    \texttt{PoWR} WNL and WNE libraries, 6 for \texttt{PoWR} WC
    library, and 10 for Koester WD library.  If a star is
    interpolated between different libraries, an intermediate number
    is assigned. However, in the future these numeric labels might be changed, if 
    more libraries are implemented or the interpolation scheme is changed.}
  \label{fig:CMD}
\end{figure*}

\subsection{Stellar spectral library selection}
The above-described libraries can be used either combined or
separately. The default library selection scheme is:
\begin{itemize}
\item [1)] \texttt{ATLAS9} is used when $\teff>6,500$K;
\item [2)] \texttt{PHOENIX} is used when $\teff<5,500$K;
\item [3)] a smooth interpolation between the previous two when $\geq
  \teff \leq 6,500$K;
\item [4)] \texttt{COMARCS} M/S star library is used when
  $(X_\mathrm{C}/12)/(X_\mathrm{O}/16)\leq1$ and $\teff<4850$K and $\logg > 1.5$;
\item [5)] a smooth interpolation between \texttt{PHOENIX} and
  \texttt{COMARCS} M/S star library in the overlapping region between
  them;
\item [6)] \texttt{COMARCS} C star library is used when
  $(X_\mathrm{C}/12)/(X_\mathrm{O}/16)>1$ and $\teff<4500$K and $\logg < 2.5$;
\item [7)] \texttt{WM-basic} libraries are used when $\teff > 20,000$K
  and the mass loss rate $\dot{M}>10^{-8}\,M_{\odot}\,\mathrm{yr}^{-1}$;
\item [8)] \texttt{PoWR} libraries are used when $\teff > 100, 000$K
  or $X < 0.8 X_\mathrm{i}$ or $X<0.65$, where $X_\mathrm{i}$ represents the original
  surface hydrogen mass fraction;
\item [9)] when \texttt{PoWR} libraries are selected and $X_\mathrm{C}/12 > X_\mathrm{N}/14$
  and $X_\mathrm{C}>0.05$, \texttt{PoWR} WC library is used;
\item [10)] Koester WD library is used when $\logg>6$ and
  $\teff>6,300$K.
\end{itemize}
The stellar surface chemical compositions are used in the above
scheme, therefore they are required in the web interface (see
Appendix~\ref{app:interface} for more details).  If the required
chemical abundances are not provided by the user, a solar scaled
abundance is used with the specified metallicity $Z$, which means all
the relevant abundance ratios are the same as those in the Sun.  As a
test case we have applied the new bolometric corrections to the PARSEC
V1.2s isochrones.  In Figure~\ref{fig:CMD}, we show the isochrones
with [M/H]=0 and for selected ages, in the standardised $JK$ system
from \citet{Bessell1990}, and in the Gaia DR2 passband as described in
\citet{Maiz2018} (the Gaia filters used in the following sections are
also from this reference, unless specified otherwise).
Different colours indicate the different
stellar spectral libraries adopted, as specified in the caption.

\begin{figure*}
  \includegraphics[width=0.495\textwidth]{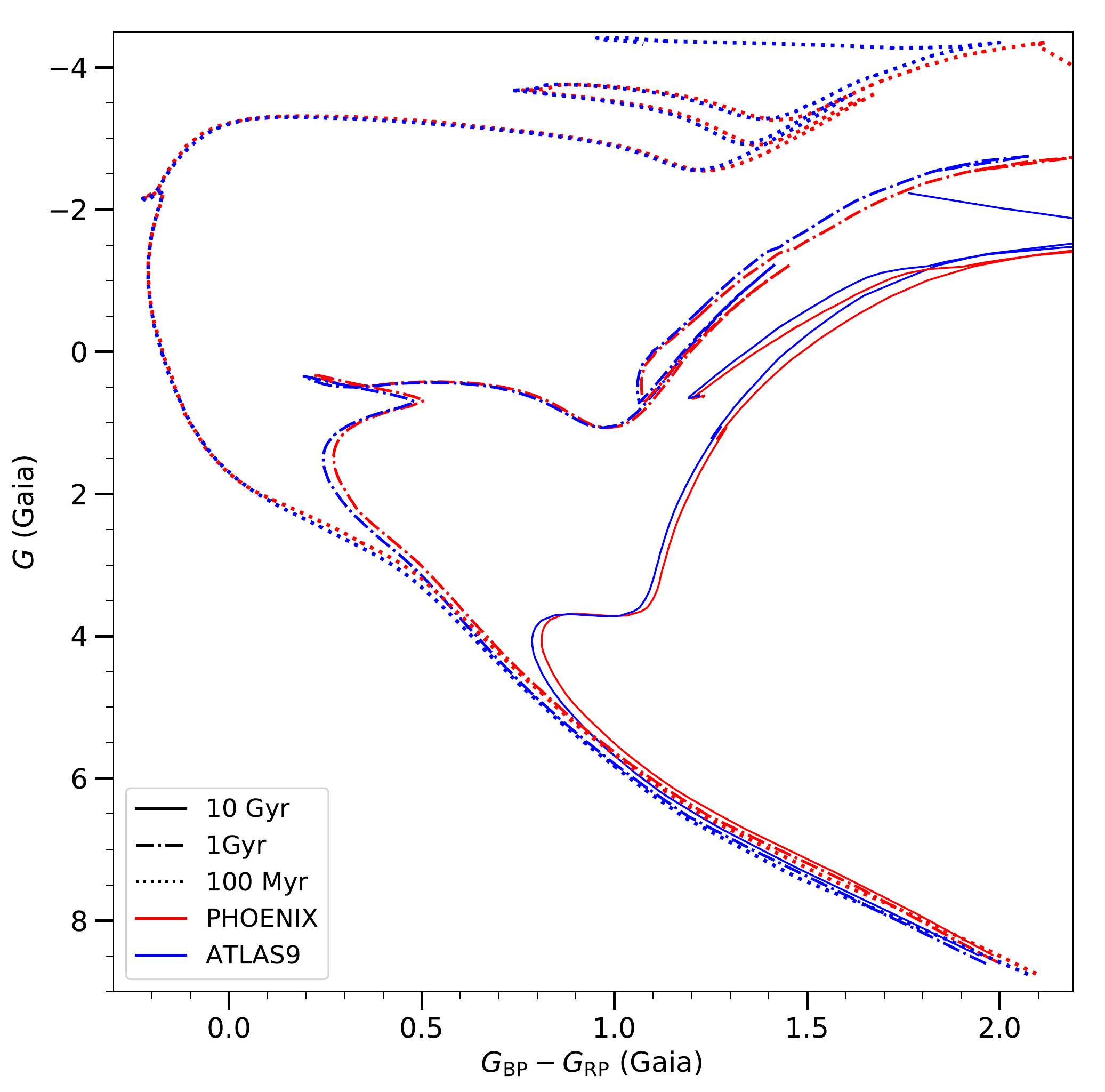}~
  \includegraphics[width=0.495\textwidth]{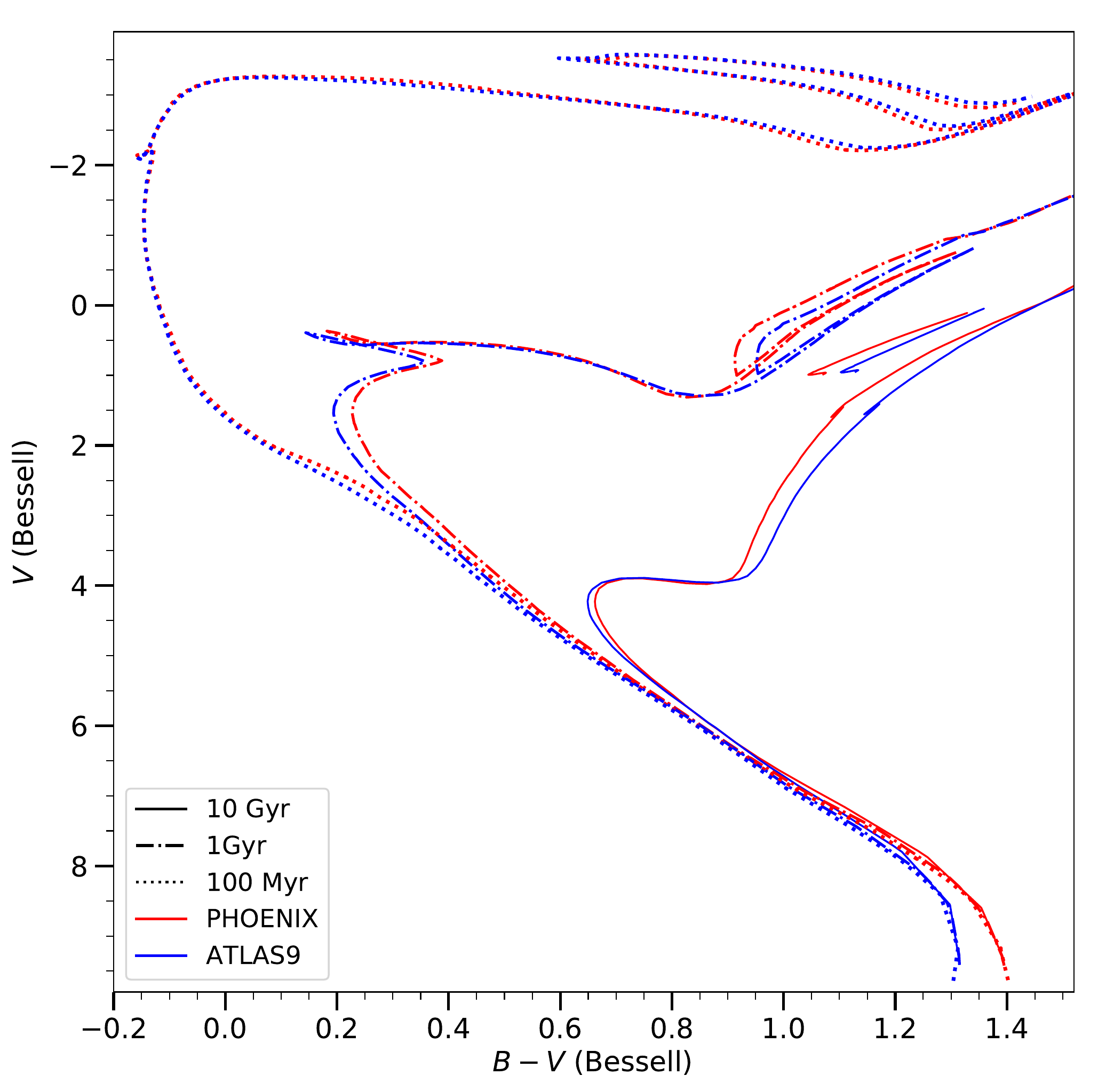}\\
  \includegraphics[width=0.495\textwidth]{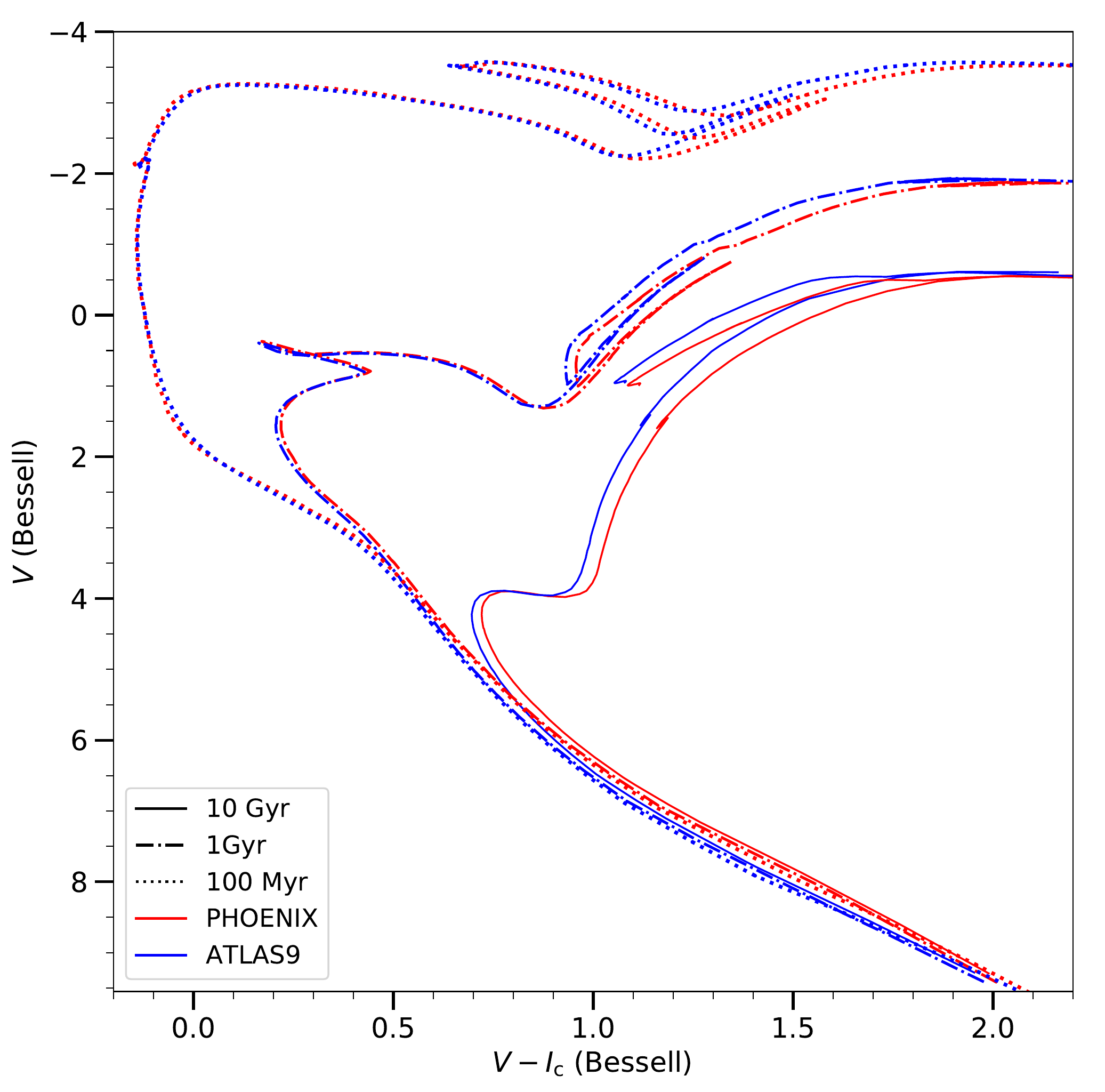}~
  \includegraphics[width=0.495\textwidth]{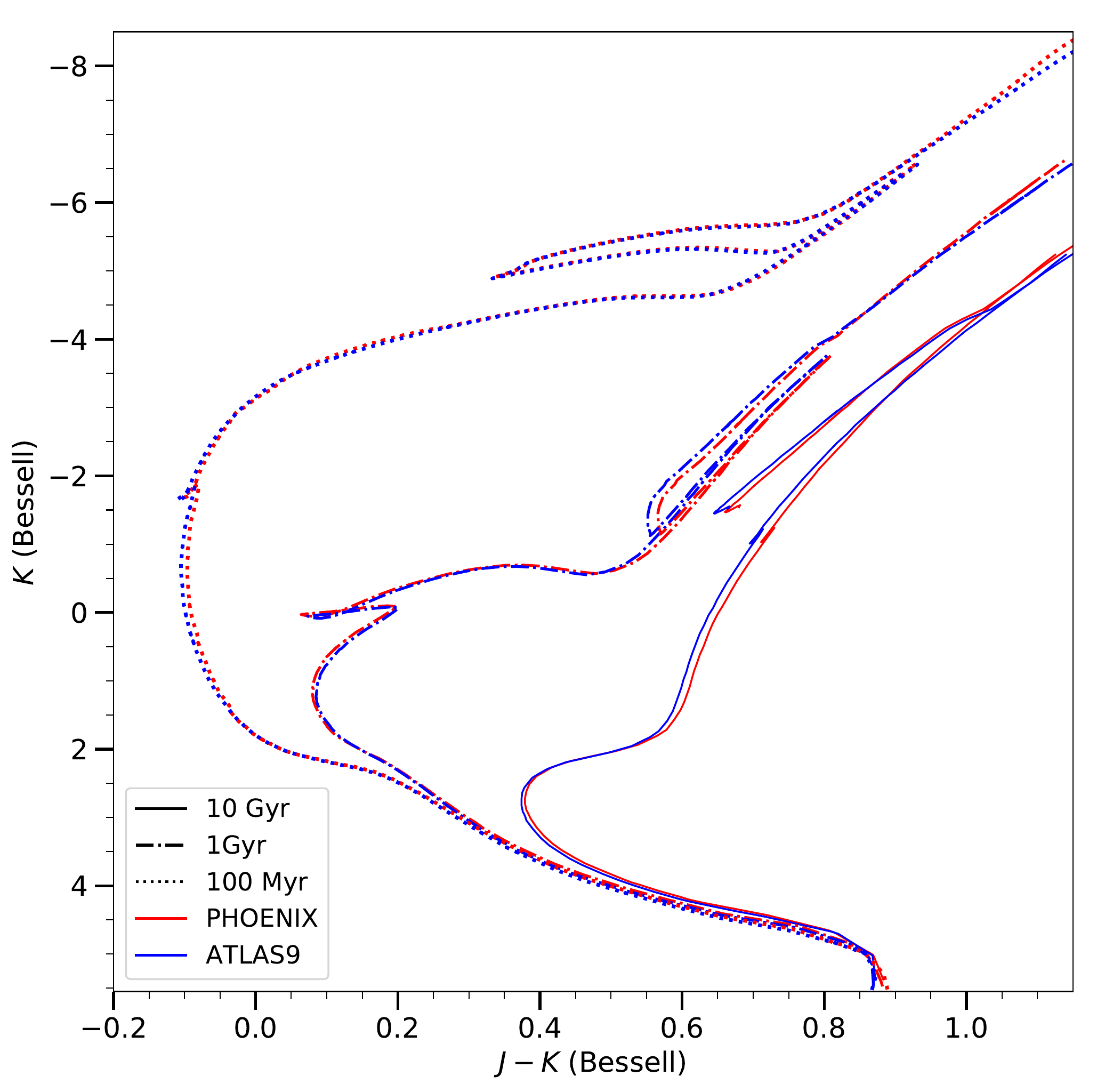}
  \caption{Differences between \texttt{PHOENIX} and \texttt{ATLAS9} models in
    the CMDs.  From left to right and top to bottom: Gaia $G_{\rm
      BP}-G_{\rm RP}$ .vs. $G$, $B-V$ .vs. $V$, $V-I_{\rm c}$ .vs. $V$ and $J-K$ .vs. $K$. 
    The Gaia filters are from \citet{Maiz2018}, while the standard filters are from \citet{Bessell1988,Bessell1990}.  
    \texttt{PARSEC} v1.2S isochrones of
    [M/H]=0.0 and ages log(age/yr)=8, 9 and 10 are plotted. The
    \texttt{ATLAS9} models are only calculated for stars hotter than
    above 3,500\,K, therefore those parts of the isochrones with
    $\teff$ below 3,500\,K are not shown.}
  \label{fig:CMD_diff}
\end{figure*}

Beside the above default scheme, we also offer some other options,
such as \texttt{ATLAS9} only, \texttt{PHOENIX} only, or a combination
of them (with an adjustable transition region), etc.  An issue
concerns the difference in broad band CMDs between different
libraries.  We take the comparison between \texttt{ATLAS9} and
\texttt{PHOENIX} as an example to address this issue.  In
Figure~\ref{fig:CMD_diff} we show the differences between
\texttt{PHOENIX} and \texttt{ATLAS9} libraries in several CMDs, using
the same \texttt{PARSEC} v1.2S isochrones of [M/H]=0.0 and ages log(age/yr)=8,
9 and 10. The difference between \texttt{PHOENIX} and \texttt{ATLAS9}
increases at decreasing $\teff$.
At the solar metallicity, the colour difference between these two
atmosphere models can reach $\sim 0.05$mag at the main sequence
turn-off and $\sim 0.1$mag at the giant branch in the V-(V-I) plane.
For the lower main sequence stars, we have shown in our previous work
\citep{Chen2014} that \texttt{PHEONIX} offers a better fitting
compared to \texttt{ATLAS9}. 
Moreover, the \texttt{PHOENIX} library is computed
with spherical geometry, making it preferable than \texttt{ATLAS9}
models especially for giants.

\section{The code package and web interface}
\label{sec:package}
The design of this package is to make it transparent to the user,
extensible to adopting new stellar spectral libraries, filter sets,
and definitions of photometric systems.

The whole package is divided into three parts/steps. In the first
step, the original stellar spectral libraries are assembled and
re-sampled to FITS/HDF5 format files. In the second step, these files are
used to compute bolometric corrections. In the third step, these BC
tables are used in interpolation for stars of given stellar
parameters. 
The code for the first and second steps is written in
Python because the computational speed is not an issue and we only need to compute those tables once for all.
Compared to C or Fortran, Python provides high-level libraries for manipulating tabular data (including ASCII, FITS and HDF5 formats).
The third step is written in C language for
a faster computational speed.

\subsection{Assembled FITS/HDF5 files of spectra}
In this step, the original spectra are assembled into FITS
\citep{FITS} and/or HDF5 \citep{HDF5} format files.  If the original
spectra are in very high resolution, eg. the \texttt{PHOENIX} models contain
$\sim23000$ wavelength points which are not necessary for the BC
calculation except for narrow band filters, they are re-sampled into a
lower resolution wavelength grid before assembling for
computational speed and file size. The re-sampling is done with a
modified version of \texttt{SpectRes} package \citep{SpectRes}, which preserves
the integrated flux within each of the wavelength grids.  After this
step, the stellar spectral libraries are store in homogeneous file
format and organisation. This step is necessary as the original
stellar spectral libraries are provided in different formats. For
example, the \texttt{PHOENIX} models are stored in a single ASCII file for each
spectrum, while \texttt{ATLAS9} models are stored in one single ASCII file
for all the models of a given metallicity. 
Currently, the assembled spectra for non-rotating stars are stored in FITS format and can be easily viewed with \texttt{TOPCAT}
\citep{topcat2017}, while the assembled spectra for rotating stars are stored in HDF5 format for their logically complexity.

\subsection{Computing tables of BCs}
In this step, the above-generated FITS format spectra are read to
compute the BCs, for different filter sets available, according to
equations (\ref{eq_BC_photon}) to (\ref{eq_BC_energy_AB}). The BCs are
first computed for the original grids. However, these original grids
are usually not rectangular or uniformly distributed in the
$\logteff$--$\logg$ ($\log R_\mathrm{t}$) plane. For example, at high
$\logteff$, usually the high $\logg$ models are missing due to the
proximity to the Eddington limit. We always extend the grids in
$\logg$ with neighbouring models of the same $\logteff$.  Thereafter,
these rectangular grids are re-sampled in $\logteff$--$\logg$ ($\log
R_\mathrm{t}$) plane with fixed steps. This re-sampling enables fast index
searching for the interpolation when utilising the BC tables. These
final BC tables are stored into FITS format files.
BCs with different $\Av$s are assembled in to the same file, therefore the
reader can easily check some of the discussions presented in section~\ref{sec:extinction}
(such as Figure~\ref{fig:BCAv}).
Furthermore, this Python package provides an option to use
``gnu-parallel'' tool \citep{Tange2011} for computing tables for many
different filter sets in parallel.

\subsection{Interpolation scheme}
After the BC tables are computed, users can use these
tables with their own interpolation code, or a C code provided by us upon request.
In summary, this code employs linear interpolations in $\logteff$, $\logg$ or ${\rm log}R_\mathrm{t}$, 
${\rm log}\dot{M}$ and ${\rm log}Z$.
First, the BC tables in a
specified filter set are loaded into the memory. An input star is then
assigned a label according to its $\logteff$ and $\logg$ (or $\log
R_\mathrm{t}$ for WRs), or also spectral type in the case of WRs. This label
tells the code which libraries are assigned. If this star is in the
transition region between two libraries, both libraries are
selected. A weight is given according to the proximity of the star to
each library. Within a library, the interpolation is done in the
$\logteff$ and $\logg$ grids of two neighbouring metallicities and
then between the two metallicities. Because the $\logteff$ and $\logg$
grids are pre-sampled at fixed grids, the interpolation in the
$\logteff$ and $\logg$ plane is very fast. For instance, with a computer of 3.6 GHz
CPU, it takes less than 20~seconds for a catalogue of $10^6$ stars in
the standard $UBVRIJHK$ filters. The memory consumption is less than
20 Mby.

In the interpolation, there are issues concerning the metallicities used in different stellar spectral libraries. 
Different libraries were computed with different solar abundances and different $\alpha$-enhancement. 
Part of the difference presented in the figure \ref{fig:CMD_diff} comes from the metallicity difference.
Moreover, the isochrones or theoretical stars may have solar abundances and $\alpha$-enhancement different from the 
stellar atmosphere models. 
Currently, we only consider the total metallicities ($Z$) for each of the libraries. The total metallicities are computed
with their own solar abundances. Though different solar abundances and $\alpha$-enhancements affect the detailed spectral 
features, the broad-band BCs are less affected and the interpolation in the total metallicities can be taken as the first-order
approximation.
We have recorded the 
solar abundances and $\alpha$-enhancements information in the code and may consider these effects when a better interpolation 
scheme is defined. 
We are also computing atmosphere models with the same abundances as used for computing \texttt{PARSEC} tracks, with the \texttt{ATLAS12} code. These models will
ensure the consistency between the atmosphere models and \texttt{PARSEC} stellar models, and may also allow us the evaluate
the above issues.

\subsection{Web interface}
We build an easy-to-use web interface
(\url{http://stev.oapd.inaf.it/YBC}) for the users to directly
convert any uploaded catalogue containing theoretical quantities into
magnitudes and colours (not necessarily the \texttt{PARSEC}
ones). More details about the web interface are provided in
  appendix~\ref{app:interface}.

The above assembled spectral libraries, as well as the BC tables in
different filters, can be downloaded at the ``Spectral libraries'' and
the ``BC tables'' sections of this website, respectively.


\begin{figure*}
  \includegraphics[trim=6 0 0 0,clip,width=0.5\textwidth]{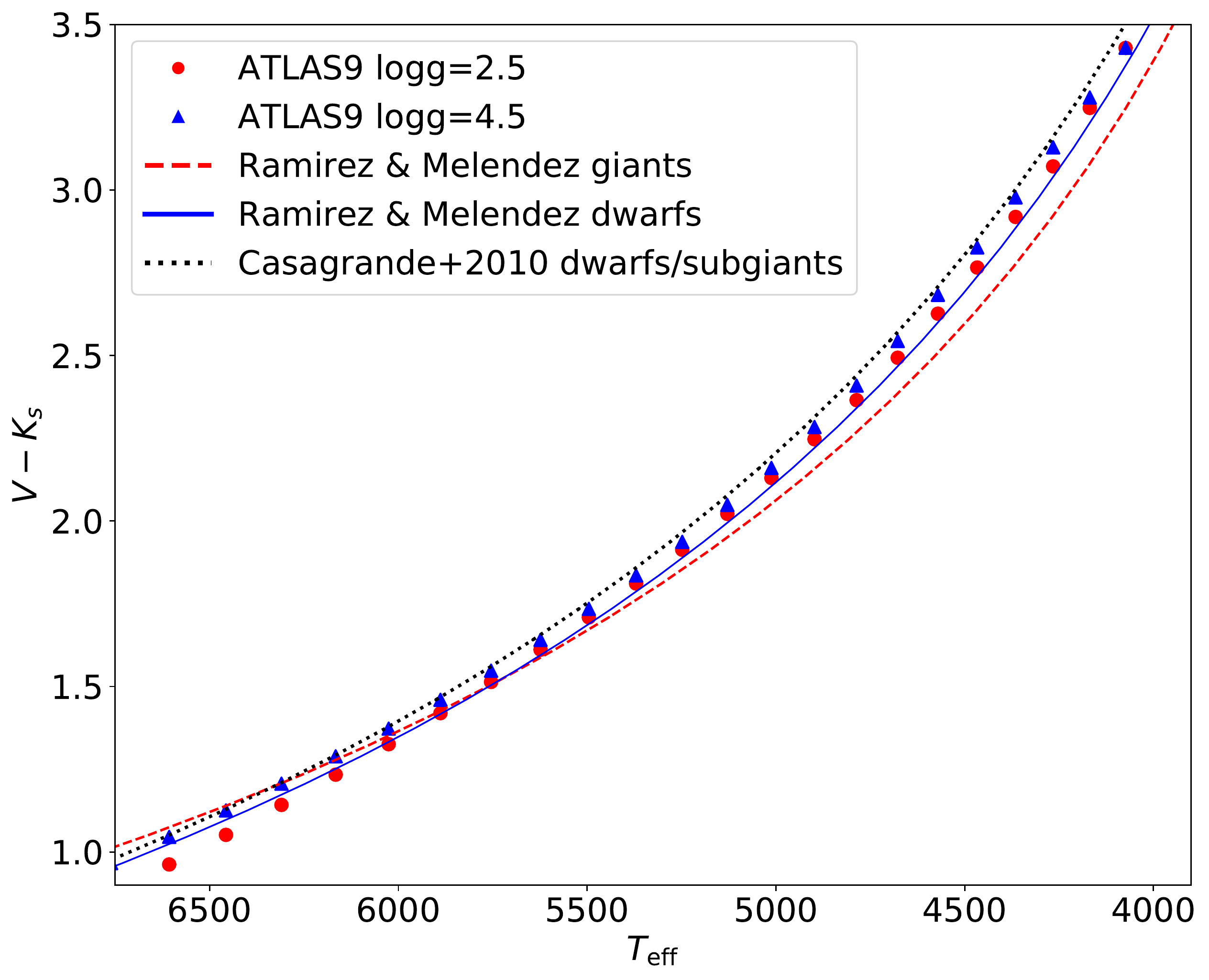}\hfill
  \includegraphics[trim=6 0 0 0,clip,width=0.5\textwidth]{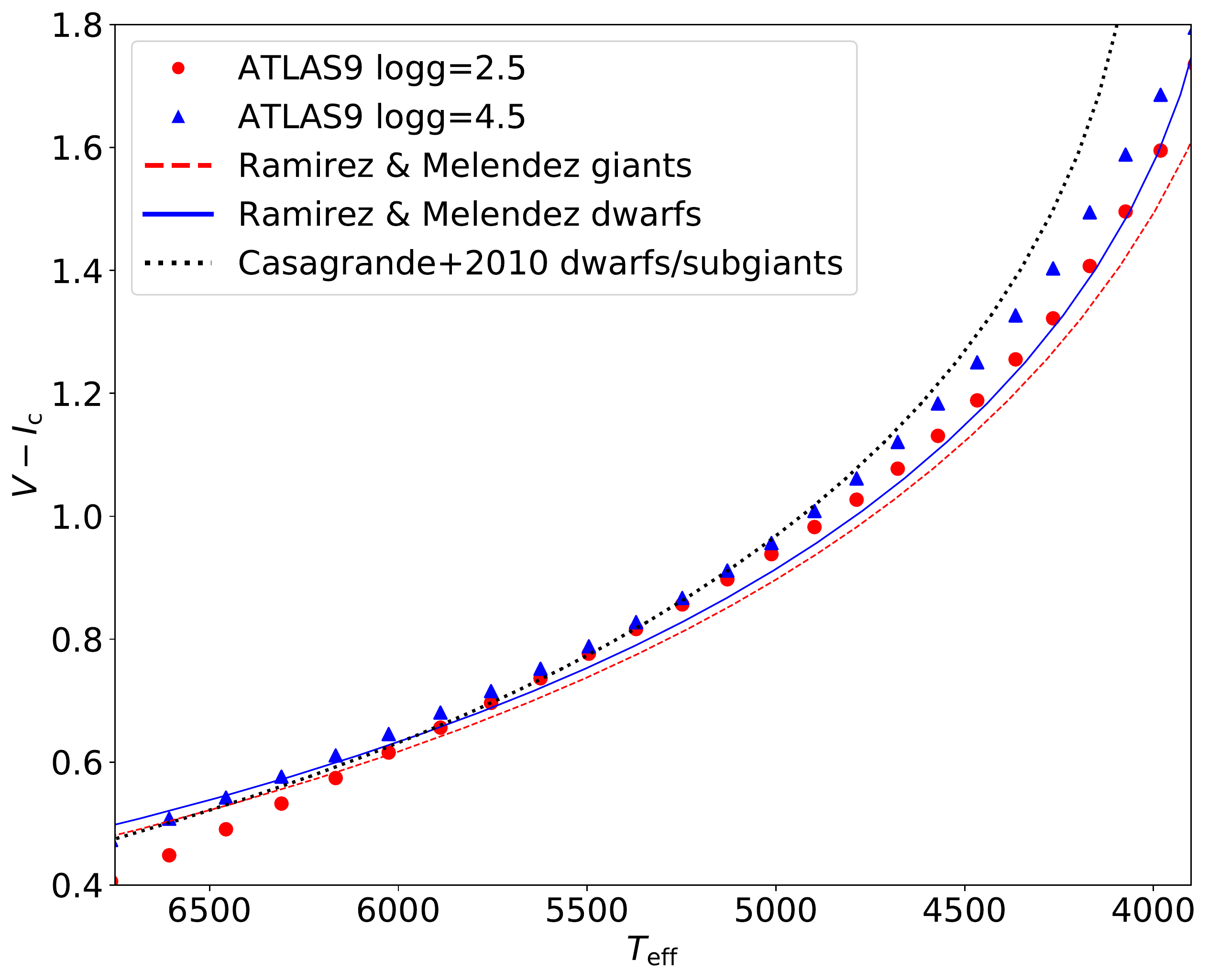}\\
  \includegraphics[trim=6 0 0 5,clip,width=0.5\textwidth]{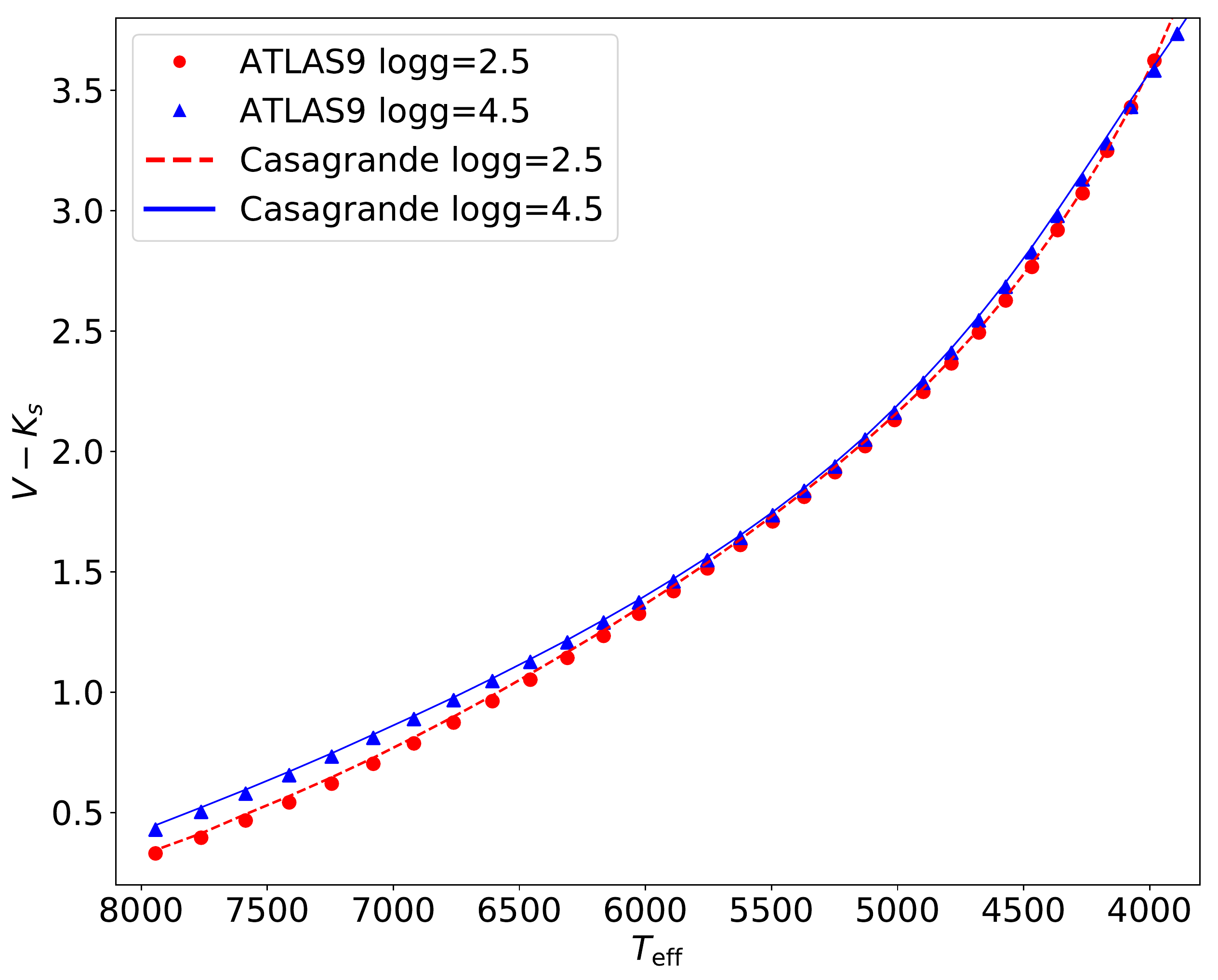}\hfill
  \includegraphics[trim=6 0 0 5,clip,width=0.5\textwidth]{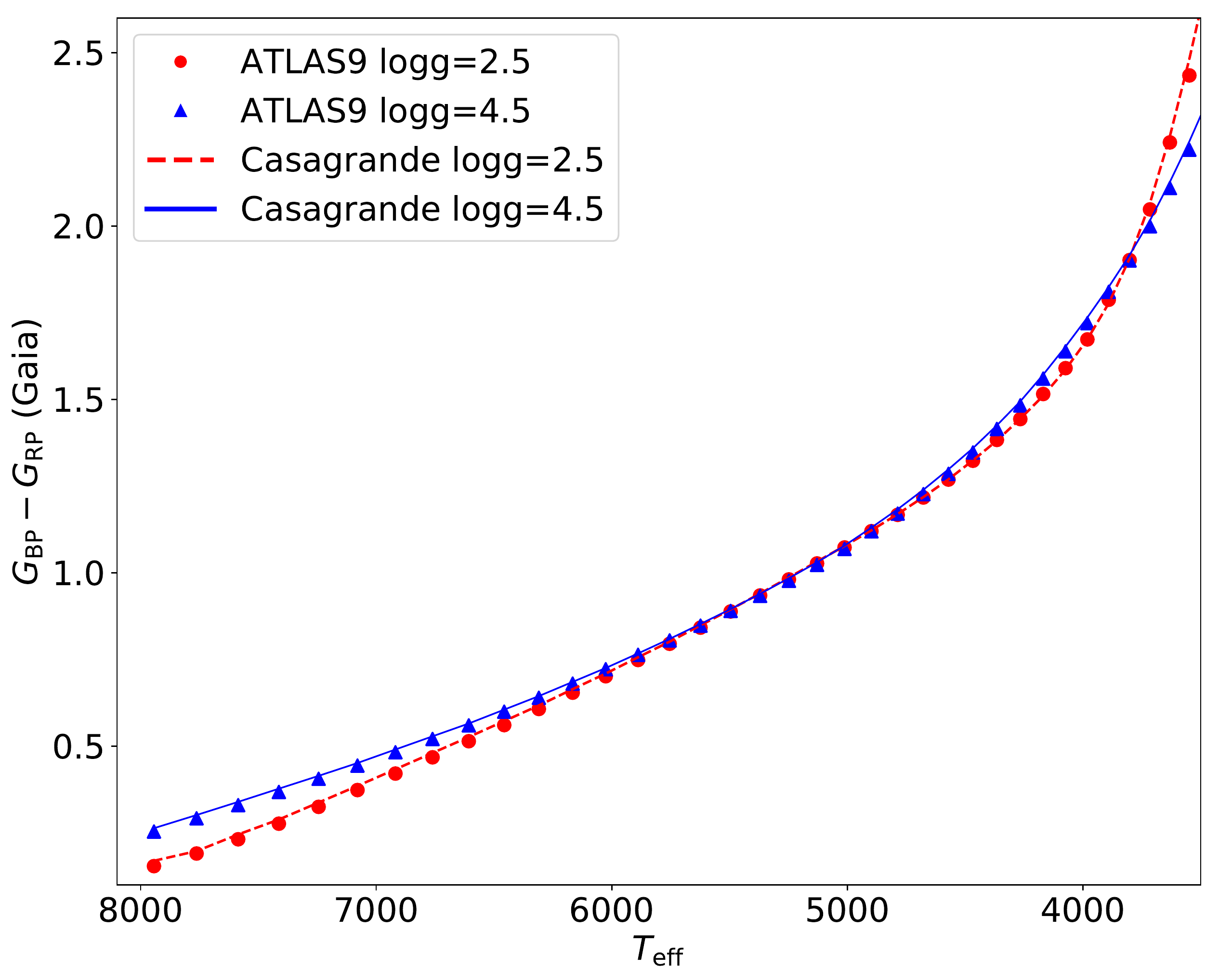}\\
  \includegraphics[trim=6 0 0 10,clip,width=0.5\textwidth]{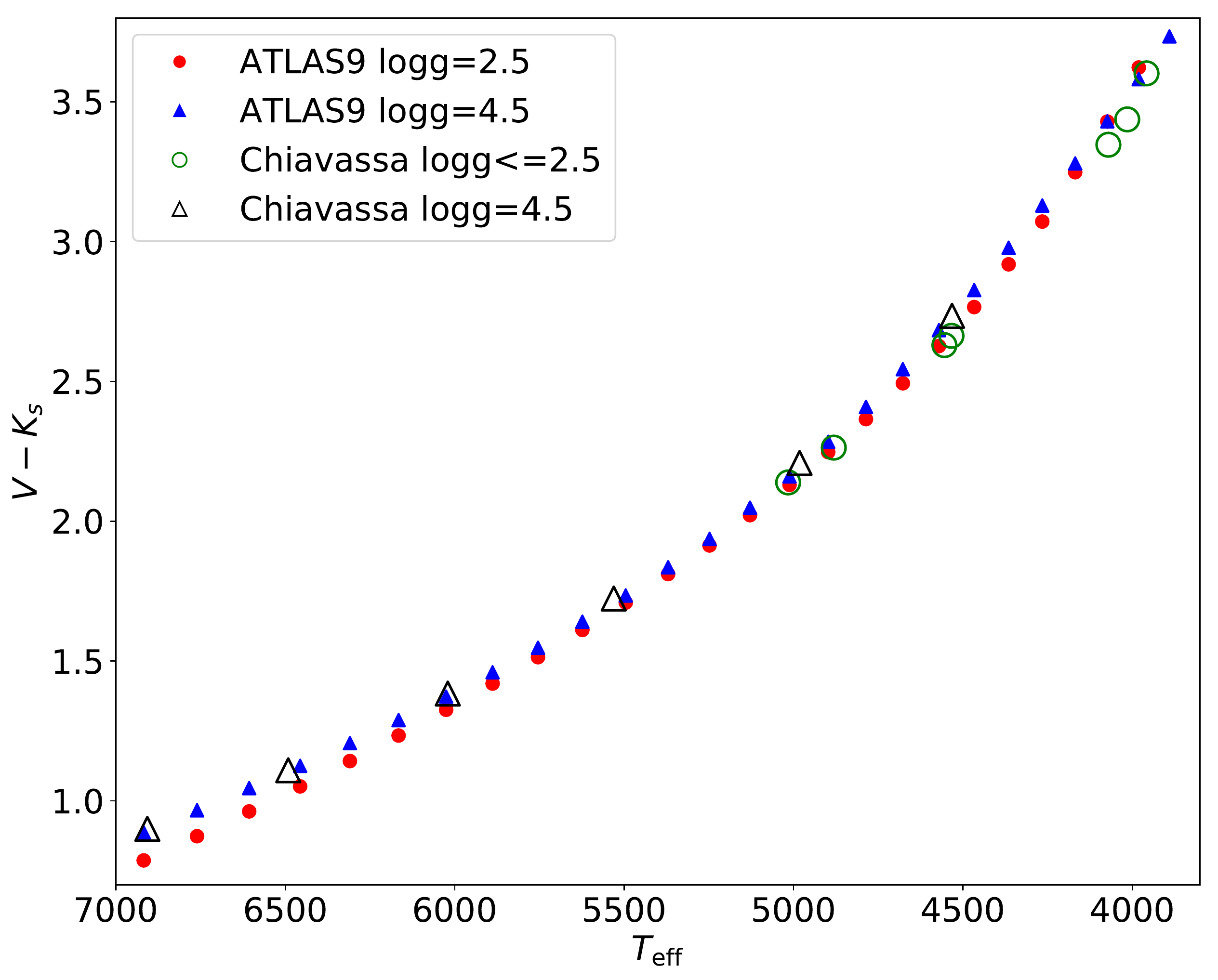}\hfill
  \includegraphics[trim=6 0 0 8,clip,width=0.5\textwidth]{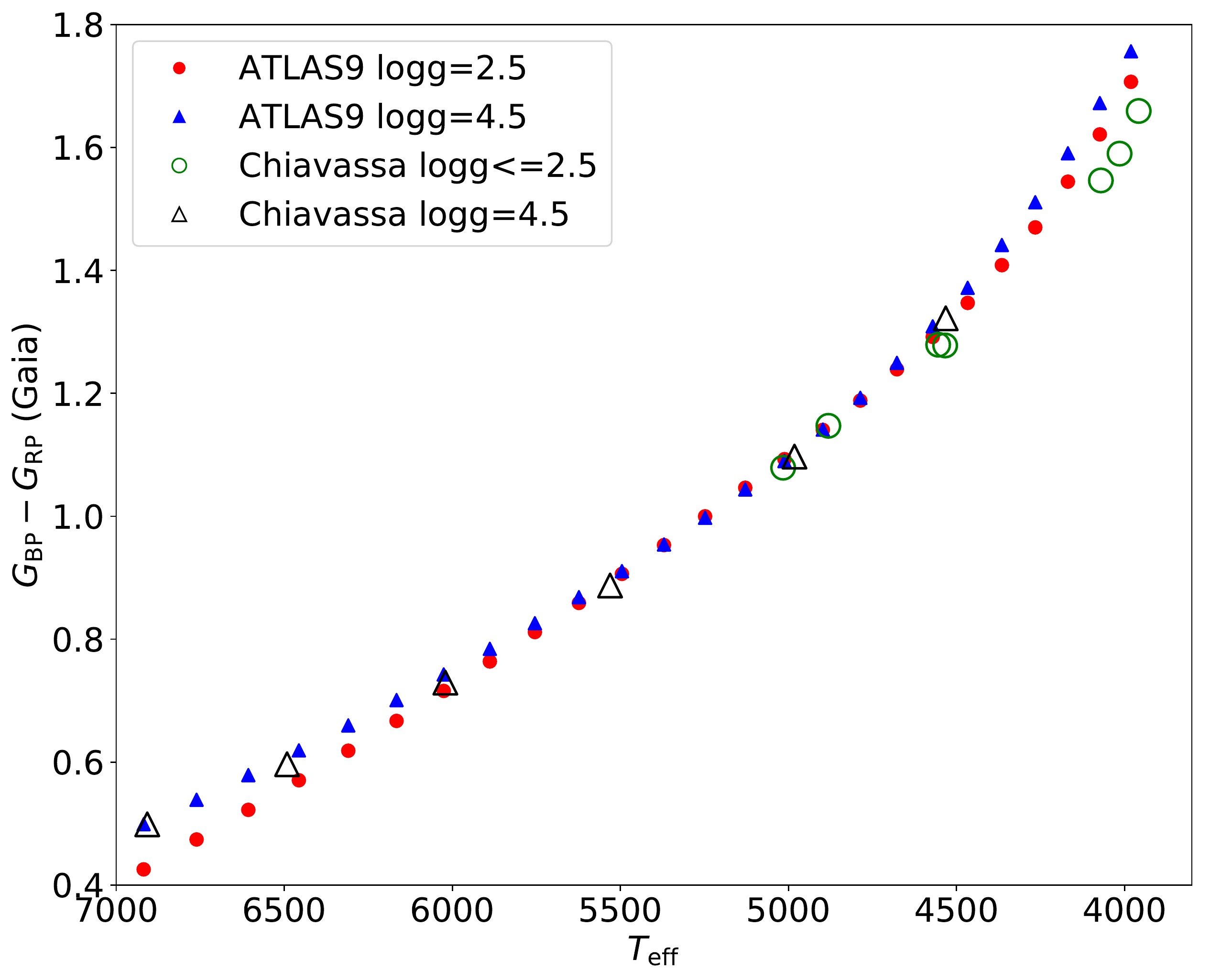}
  \caption{Comparison of the temperature scales for $\feh=0$ for the standard filters \citep{Bessell1990,Maiz2006}, 2MASS \citep{2MASS} and Gaia filters. 
  Upper panels: comparison with the empirical temperature scales from \citet{Ramirez2005} (blue solid and red dashed lines) and \citet{Casagrande2010} (black dotted lines).
  Middle panels: comparison with the results from \citet{Casagrande2014, Casagrande2018a, Casagrande2018b},
  the Gaia filters are from \citet{evans18}.
  Lower panels: comparison with the 3D models from \citet{Chiavassa2018}, the Gaia filters are those from \citet{Jordi2010}.
  Our BCs are derived from \texttt{ATLAS9} models.
  }
\label{fig:BCcomparison}
\end{figure*}

\section{Comparison with literature results}
\label{sec:bccomparison}
In this section, we compare our BCs derived from \texttt{ATLAS9} models with those from some well-known studies on the temperature scales or bolometric corrections. It is intended mainly as a general consistency check, rather than a detailed comparison.

\citet{Ramirez2005} provide empirical temperature scales for stars with $4000 \lesssim \teff \lesssim 7000$\,K and $-3.5 \lesssim {\rm [Fe/H]} \lesssim 0.4$. In the upper panels of Figure~\ref{fig:BCcomparison}, we compare their $V-K_\mathrm{s}$ and $V-I_\mathrm{c}$ vs. $\teff$ relations with ours. They are consistent to within $\sim0.1$\,mag. We also plot the empirical relations for the dwarfs/sub-giants from \citet{Casagrande2010}, which improve the comparison with the theoretical ones for the relatively hot stars, but behave oppositely for cooler stars. This may indicate that there are still some large uncertainties in determining the physical parameters for cool stars. 
\citet[][denoted as Casagrande in the figure]{Casagrande2014, Casagrande2018a, Casagrande2018b} presented a code to compute the BCs based on the \texttt{MARCS} \citep{MARCS} atmosphere models. In the middle panels of Figure~\ref{fig:BCcomparison}, we show the comparison for both the $V-K_\mathrm{s}$ and $G_\mathrm{BP}-G_\mathrm{RP}$ colours.  
We find a very good agreement between ours and theirs. 
Finally, in the bottom panels we compare our BCs with those based on 3D atmosphere models from \citet{Chiavassa2018}. The agreement is also quite reasonable, except from some departure at low $\teff$. 
This might implicate that the atmospheres of cool stars could be better modelled with 3D models.

The above comparisons indicate that there is a general agreement between theoretical BCs derived from different 1D models or 3D models (except for cool stars), while there is some non-negligible difference between 1D models and empirical relations or 3D ones, especially for cool stars, which deserves further investigation to solve the discrepancies.

\begin{figure*}
  \includegraphics[trim=8 5 5 0,clip,width=0.33\textwidth]{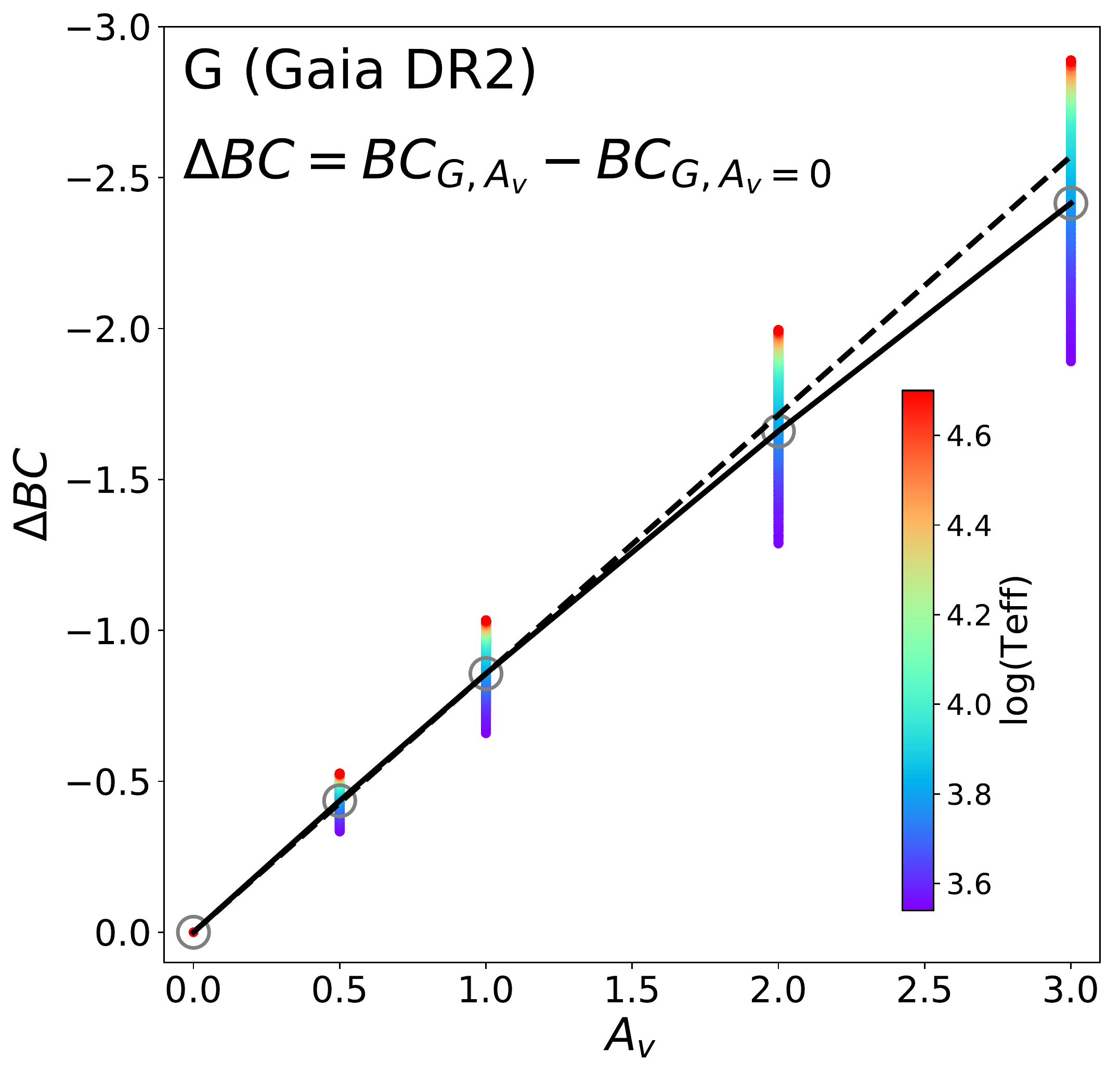}
  \includegraphics[trim=8 5 5 0,clip,width=0.33\textwidth]{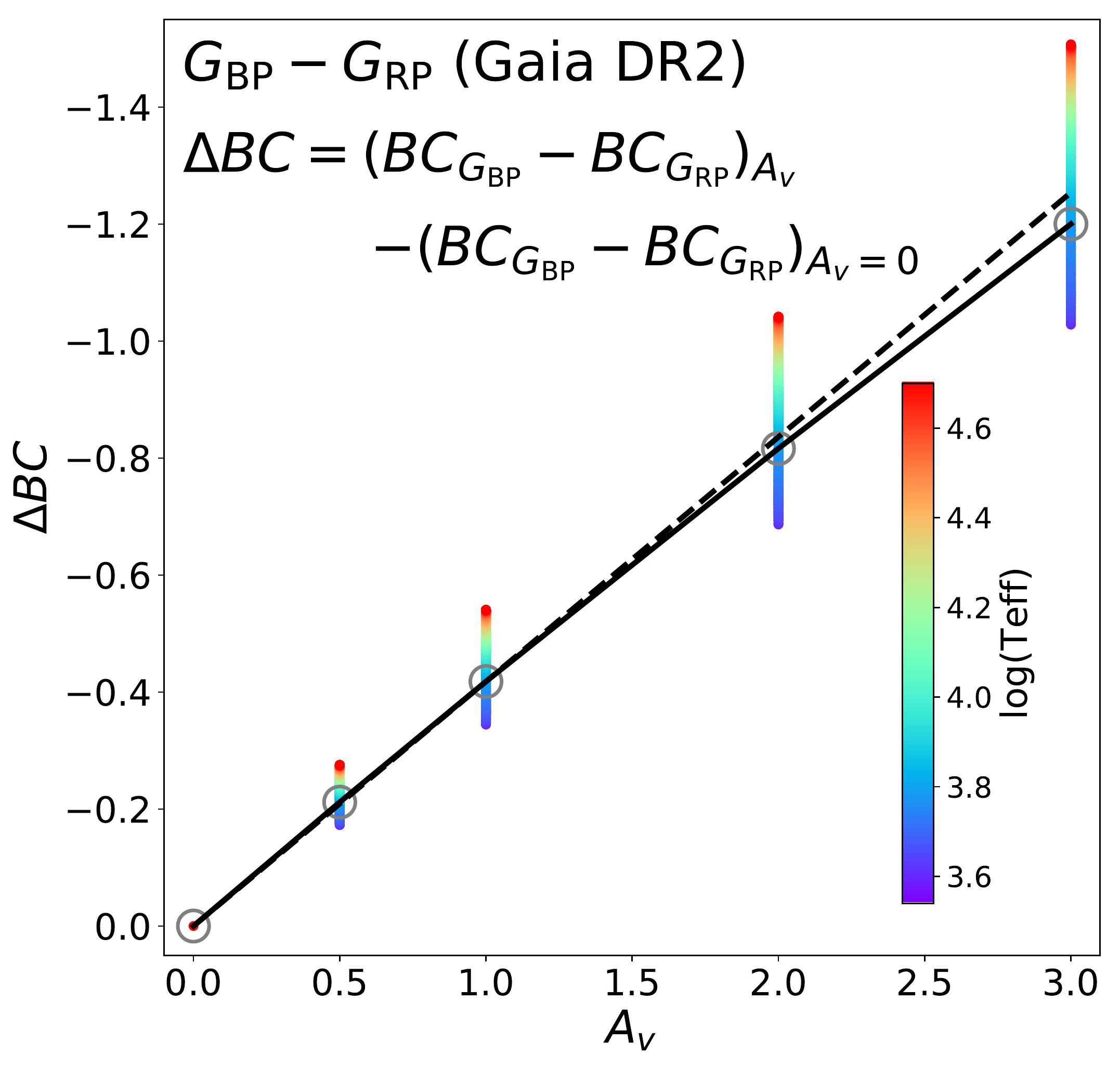}
  \includegraphics[trim=8 5 8 0,clip,width=0.32\textwidth]{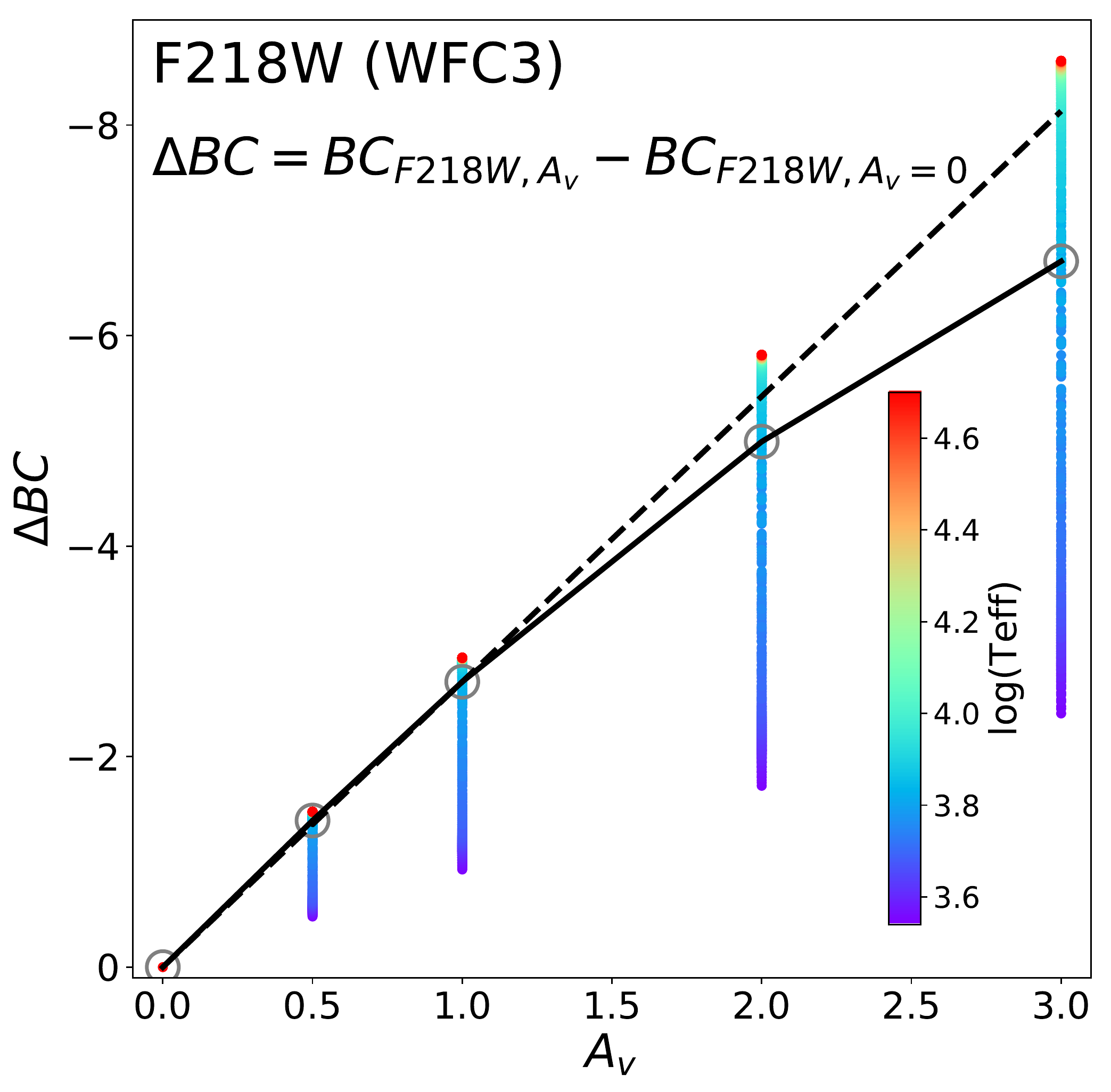}
  \caption{$\Delta BC=\BC_{\Av}-\BC_{\Av=0}$ as a function of $\teff$ 
    for Gaia $G$ band (left panel), Gaia $G_\mathrm{BP}-G_\mathrm{RP}$
    colour (central panel) and the WFC3/UVIS F218W filter (right
    panel). The grey open circles denote the models with fixed
    extinction coefficients at the filter centres, which are connected
    by the solid lines. The dashed lines are a linear extrapolation of
    the open circles of $\Av=0$ and $\Av=1$~mag, for the same
    $\teff$. The \texttt{ATLAS9} models are used here, but the
    \texttt{PHOENIX} models give consistent results.}
\label{fig:BCAv}
\end{figure*}

\section{Spectral type dependent extinction coefficients}
\label{sec:extinction}

Interstellar extinction is included into equations
(\ref{eq_BC_photon}) and (\ref{eq_BC_energy}) using extinction curves computed with the
``extinction''\footnote{\url{https://extinction.readthedocs.io}.}
Python routine, which includes most of the popular extinction laws in
the literature. By default, our database is computed using
\citet{Cardelli1989} plus \citet{O'Donnell1994} extinction law
(CCM+O94) with $R_V=3.1$, to keep consistency with the CMD web-page.
However, other extinction laws with different $R_V$ can be computed
upon request. For example, we also include the extinction curve from
\citet{Wang2019} (private communication), in which they found a
non-negligible discrepancy with that of \citet{Cardelli1989} along our
Galaxy by using a large data-set of photometry from Gaia DR2 and other
surveys. In the following discussion, if not specified,
CCM+O94 is used.

The effect of extinction on each
passband can be easily evaluated by looking at the quantity
$\Delta BC=\BC_{\Av}-\BC_{\Av=0}$. 
For very narrow filters or for hypothetical monochromatic sources, the
relationship between $\Delta BC$ and $A_V$ would be linear. However,
in the general case of a broad filter, where the stellar flux can
change significantly within the filter wavelength range, the quantity
$\Delta BC$ will vary both as a function of spectral type, and as a
function of the total $A_V$. This can be seen in
Figure~\ref{fig:BCAv}, where we plot $\Delta BC$ as a function of
$\Av$ (=1, 2, and 3~mag) for the Gaia $G$ band \citep{Maiz2018}, the Gaia
$G_\mathrm{BP}-G_\mathrm{RP}$ colour and the WFC3/UVIS F218W magnitude
\citep[also see][for ACS filters]{Girardi2008}, using the
\texttt{ATLAS9} spectral library. The grey open circles denote the
models with fixed extinction coefficients at the filter centres, which
are connected by the solid lines. At every $\Av$ the dispersion is
caused by different $\teff$ and to a lesser extent also by $\logg$
(which cannot be seen in this figure, but is clear when checking the
BC tables).
At $\Av=0.5$~mag, there is already a $\sim0.2$~mag difference
in the Gaia $G$ band, $\sim0.1$~mag in the Gaia
$G_\mathrm{BP}-G_\mathrm{RP}$ colour, and $\sim1$~mag in the F218W
filter. Therefore, we suggest to use spectral-type dependent
extinction coefficients for Gaia filters and UV filters whenever
$\Av\gtrsim 0.5$\,mag.  This dispersion increases significantly with
increasing $\Av$. At $\Av=3$~mag, for instance, the dispersion is
about 1 magnitude for the Gaia $G$ band. This means that spectral type
dependent extinction coefficients are quite necessary, especially at
large $\Av$.
Qualitatively speaking, fixed extinction coefficients would
make hot stars bluer and cool stars redder compared to the
case with variable extinction coefficients.

Furthermore, for a given spectrum and filter band,
$\Delta BC$ is not a linear function of $\Av$.
Figure~\ref{fig:BCAv} shows that at $\Av=3$ the effect brought by the
non-linearity is $\sim0.25$~mag in the Gaia
$G_\mathrm{BP}-G_\mathrm{RP}$ colour for a solar type model.
Therefore, a constant extinction coefficient (which could be derived
by only computing the BCs with $\Av=0$ and $\Av=1$~mag) for all $\Av$s is
not applicable. To properly consider the effect of extinction, we
compute the BC tables with $\Av=$[0, 0.5, 1, 2, 5, 10, 20]~mag for
each of the spectra in our database. These tables will be used for
interpolation in $\Av$, to derive BCs for any intermediate value of
$\Av$.

\begin{figure}
  \includegraphics[width=0.5\textwidth]{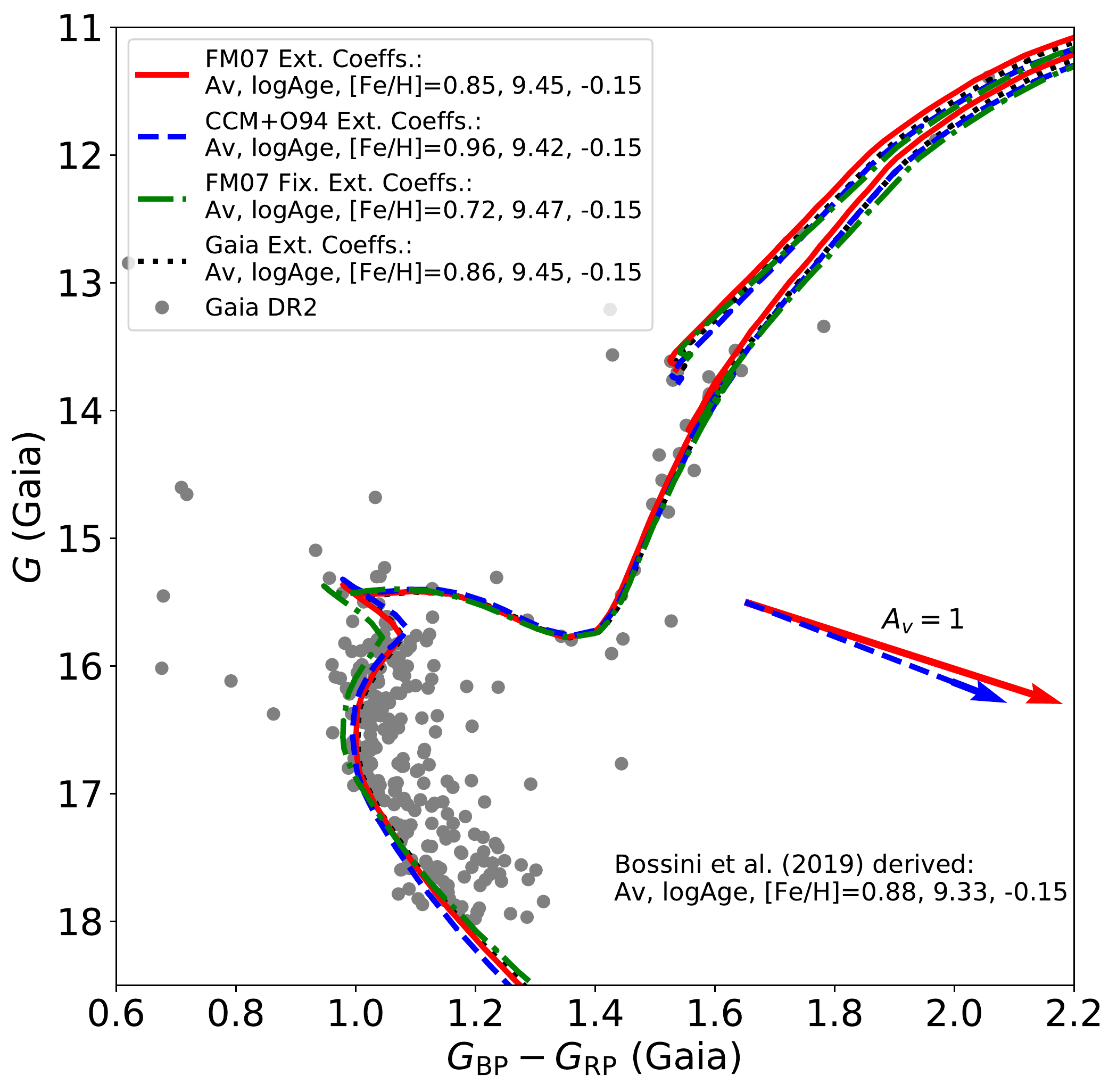}
  \caption{Isochrone fitting for NGC\,2425 with Gaia DR2 data.
    \texttt{PHOENIX} models, \texttt{ATLAS9} models and a smooth
    transition between the two are used for
    $(G_\mathrm{BP}-G_\mathrm{RP})_0 \gtrsim 0.94$,
    $(G_\mathrm{BP}-G_\mathrm{RP})_0 \lesssim 0.55$, and $0.94 \gtrsim
    (G_\mathrm{BP}-G_\mathrm{RP})_0 \gtrsim 0.55$, respectively.  For
    simplification, 
    we fix the metallicity ($\feh=-0.15$~dex)
    and the true distance modulus (12.644~mag)
    as that derived from
    \citet{Bossini2019} for all the isochrones. 
    The red solid line and
    blue dashed line are best fitting isochrones with 
    variable extinction coefficients based on FM07 and
    CCM+O94 extinction laws,
    respectively. The vectors represent the corresponding fixed
    extinction coefficients.
    Green dot-dashed line is the besting fitting isochrone with fixed
    extinction coefficients based on FM07 extinction law.
    Black dotted line
    shows the isochrone of the same age and metallicity as that of the
    red solid line, but with the extinction coefficients from that
    of \citet{Bossini2019}.
  }
  \label{fig:NGC2425}
\end{figure}

Here we show an example of isochrone fitting for NGC\,2425 with Gaia
DR2 data.
The cluster is chosen because of its relatively
large extinction and relatively old age.
For this open cluster, \citet{Bossini2019}
presented isochrone fitting parameters of $\Av=0.88$~mag, a true distance modulus 
DM$_0$=12.644~mag, age=2.15\,Gyr and $\feh=-0.15$~dex. Our intention
here is just to demonstrate the effect of variable extinction
coefficients, rather than obtaining a perfect isochrone
fitting. Therefore, we fix $\feh=-0.15$~dex and the true distance modulus (12.644~mag) as
in \citet{Bossini2019}, and vary only the way the extinction is
applied to isochrones, and their ages. 
The best-fitting isochrones with variable extinction coefficients based
on \citet[][FM07]{FM07} and CCM+O94 extinction laws are plotted with red solid
and blue dashed lines. The isochrone with FM07 extinction law has an 0.23\,Gyr older age
and 0.1 mag smaller extinction than that with CCM+O94 extinction law.
The best-fitting isochrones with fixed extinction coefficients based on FM07 extinction law
is plotted as the green dot-dashed line.
It has an 0.16\,Gyr older age and 0.25\,mag smaller extinction.
These numbers represent the uncertainties in isochrone fitting for
an object with $\Av\sim1$~mag, when different extinction approaches are used. 
We also notice that our best fitting isochrone with
FM07 extinction law 
predicts $\sim1.25$\,Gyr older age but similar extinction than
that of \citet{Bossini2019}, though we used the same \texttt{PARSEC}
model set, atmosphere models. The extinction coefficients used in \citet{Bossini2019} (which is from \citet{Babusiaux2018}) are
derived with FM07 extinction law. We plot the same isochrone as the red solid line but with
the extinction coefficients from \citet{Bossini2019}, which is shown as the black dotted line.
They are quite similar, 
thus the large difference in age
should be due to other sources. However, this is out of the scope of this work.

\begin{figure*}
  \includegraphics[width=0.495\textwidth]{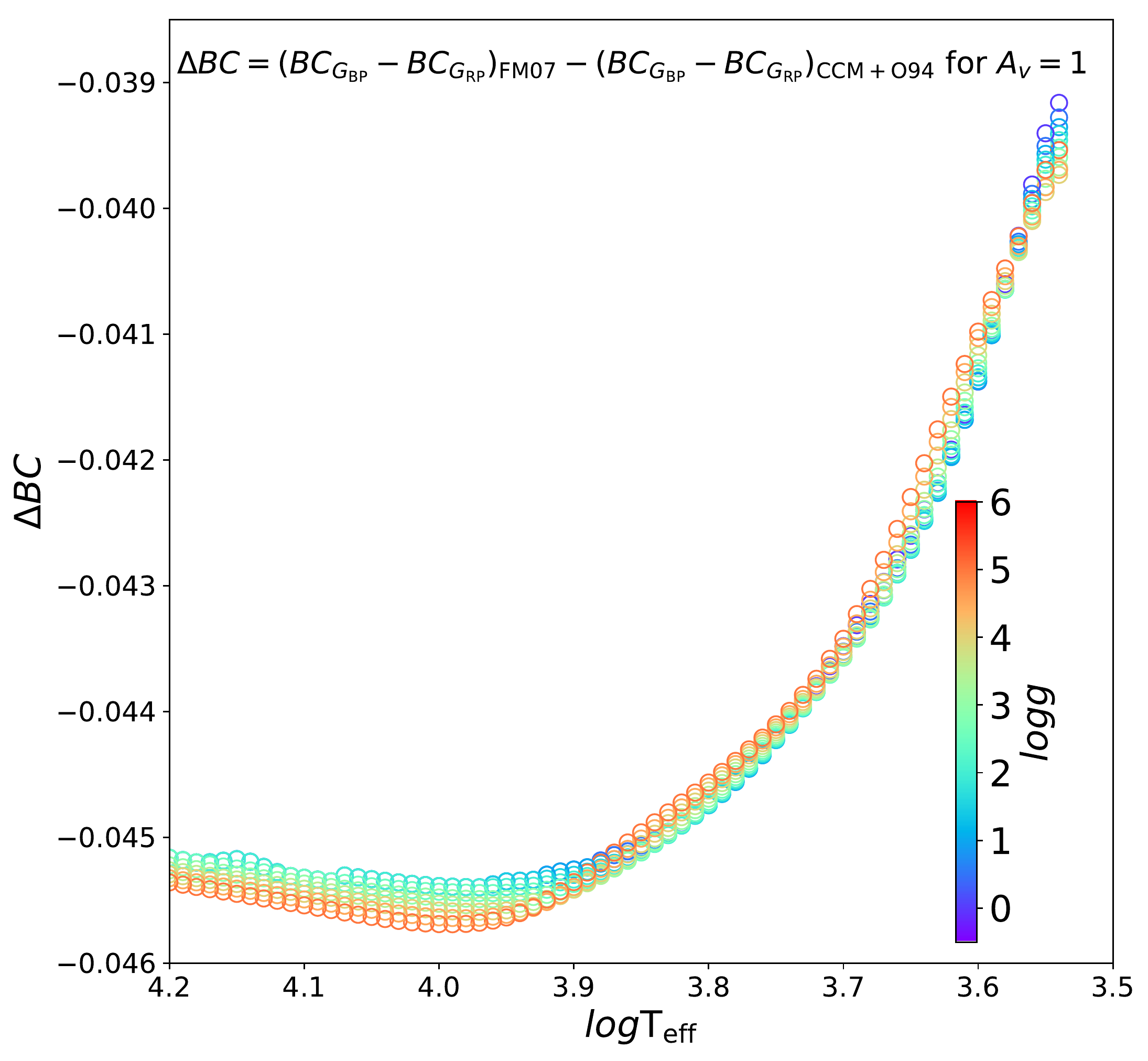}\hfill
  \includegraphics[width=0.49\textwidth]{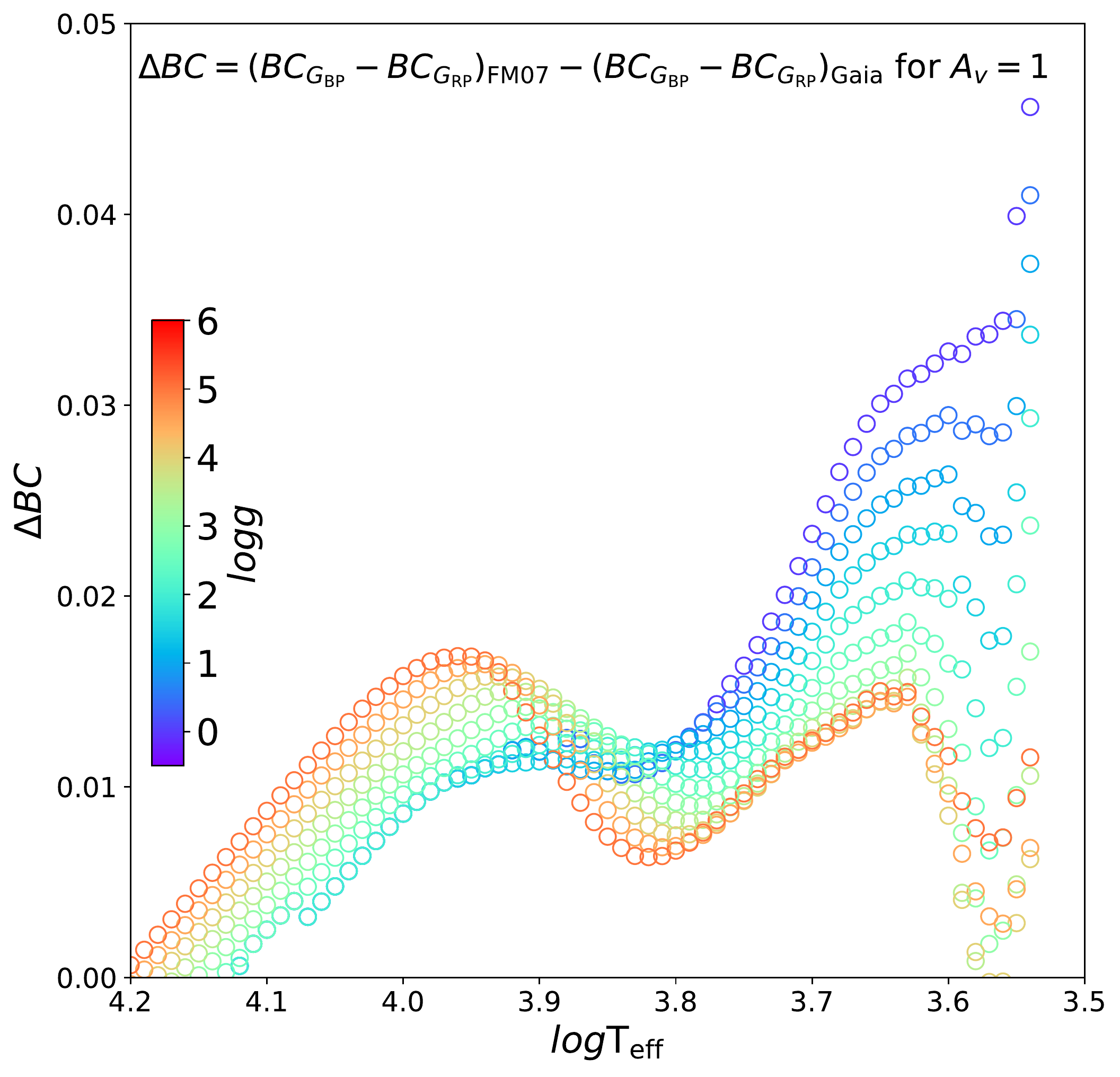}
  \caption{Differences of $\BC_{G_\mathrm{BP}}-\BC_{G_\mathrm{RP}}$ for
    $\Av=1$ using different extinction laws as a
    function of $\teff$ and $\logg$.
    Left: differences between using FM07 and CCM+O94 extinction laws.
    Right: differences between using our extinction coefficients
    based on FM07 extinction law and that of \citet{Bossini2019}.
    The \texttt{ATLAS9} models are
    used here, but the \texttt{PHOENIX} models give consistent
    results.}
  \label{fig:Ext_coeff_diff}
\end{figure*}

To further investigate the effect of using different extinction law,
we plot the BC differences with FM07, CCM+O94 and that of \citet{Bossini2019} in
Figure~\ref{fig:Ext_coeff_diff}. In the range of
$\logteff\sim3.5-3.8$, which comprises most of the NGC~2425 data, the
colour difference between FM07 and CCM+O94 cases is about 0.41 mag, and it explains most of the
differences in the derived physical parameters for isochrone fitting.
the colour difference between FM07 and that of \citet{Bossini2019} cases is about 0.15 mag,
to which some numerical errors might have contributed.
Furthermore, at lower $\logteff$, there is a quite large dispersion in
the BCs due to the variations in $\logg$.
We thus find that, for the stellar atmospheres adopted here,
the effect of using different extinction law and the gravity dependency at low temperatures is not small and should be
included in extinction parameterisation.

The above discussions imply that to derive the stellar/cluster
age and extinction more accurately, BCs that take into account the
spectral type are urgently needed, especially for objects with large extinction.

\section{Summary and Conclusions}
\label{sec:discussion}
In this paper, we present a homogeneous database of bolometric
corrections (\texttt{YBC}) for a large number of popular photometry
systems, based on a variety of stellar spectral libraries we collected
from the public domains, or computed in our previous works. In this
database, the BC tables both without extinction and with extinctions
until $\Av=20$ are computed. Therefore, our \texttt{YBC} database
provides a more realistic way to fit isochrones with a spectral type
dependent extinction, and allow users to choose the atmosphere models
that best for their science needs. The YBC database and software
package are incorporated into the PARSEC isochrones provided via the
web interface
CMD\footnote{\url{http://stev.oapd.inaf.it/cgi-bin/cmd}.}.

A potential application of this database is to incorporate it into
large simulation programs, with the interpolation routine
provided. For example, we have implemented this database in to
\texttt{TRILEGAL} for stellar population simulations \citep{trilegal2005,trilegal2016}.

Our database
can be also quite useful for stellar evolution model comparison. For
example, people studying star clusters may fit stellar evolution
models from different groups to the observation data. However,
different groups may provide photometric magnitudes with BCs from
different libraries. Through our package, they can convert theoretical
quantities into photometric magnitudes of the same libraries.

\begin{figure}
  \includegraphics[width=0.5\textwidth]{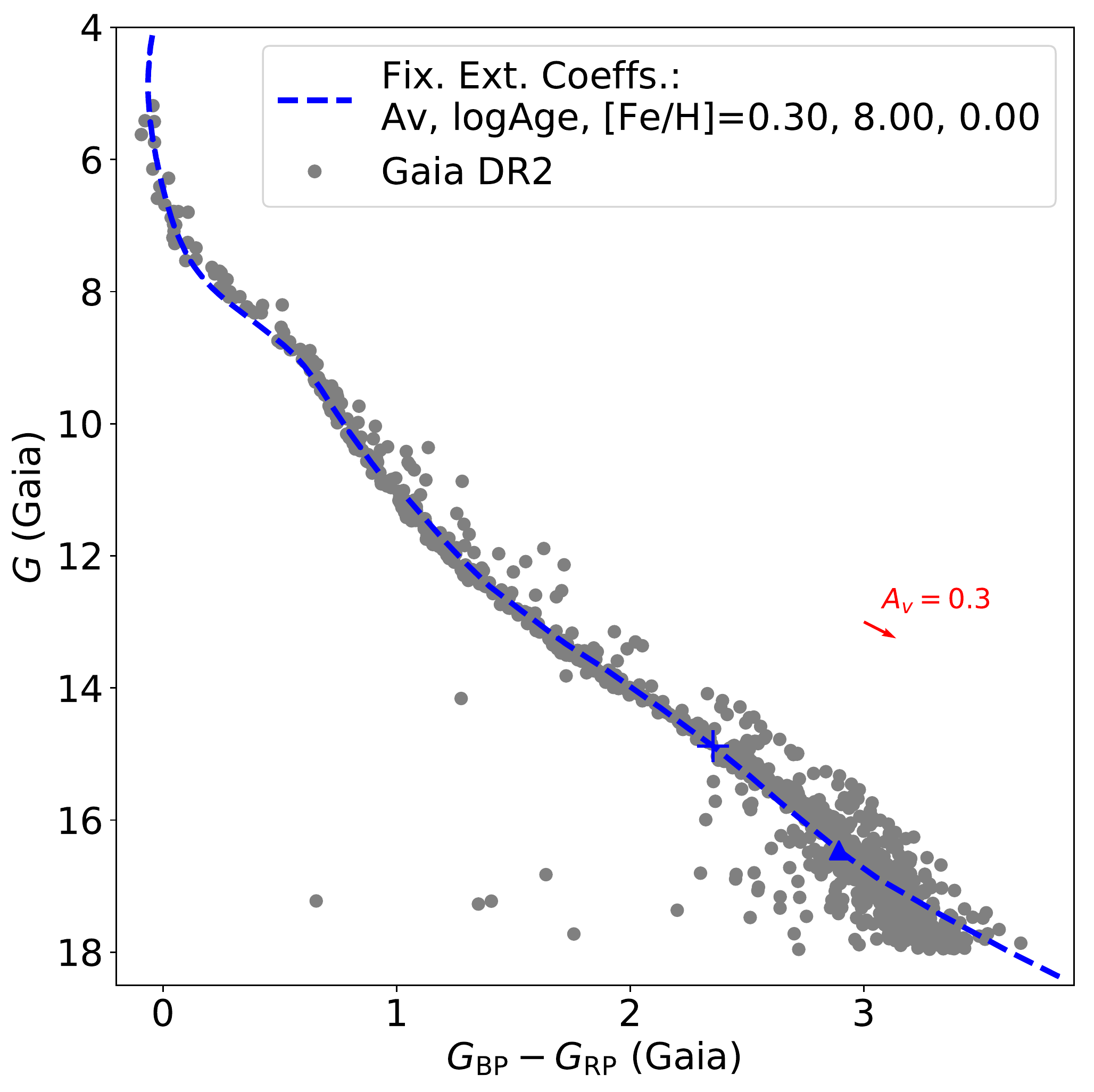}
  \caption{Isochrone fitting for Melotte~22 with Gaia DR2 data.
    The atmosphere models are the same as that in Figure \ref{fig:NGC2425}.
    We fix the metallicity ([Fe/H]=0.0)
    and the true distance modulus (5.667~mag) 
    as that derived from
    \citet{Bossini2019} for all the isochrones.
    The red cross and
    triangle represent stars of initial masses of 0.5 and 0.3 \Msun,
    respectively. For this cluster, the extinction is quite small, so
    that isochrones with both fixed and variable extinction
    coefficients do not differ much.}
  \label{fig:Melotte22}
\end{figure}

Finally, we also show an example of the isochrone fitting to a younger
cluster, Melotte~22, with the Gaia DR2 data in
Fig.~\ref{fig:Melotte22}.  It presents a quite well sampled
main-sequence, spanning almost 3.5 magnitudes in
$G_\mathrm{BP}-G_\mathrm{RP}$ colour. It can be seen that isochrones
adopting the present BC tables describe very well the entire sequence,
except perhaps for the reddest and faintest stars, which can be
affected by larger photometric errors, as well as by uncertainties in pre-MS models and in their surface boundary conditions
\citep[see][]{Chen2015}.

Work is ongoing to compute $\alpha$-enhanced evolutionary models with
and without rotation with the \texttt{PARSEC} code.  Meanwhile, we are
also computing $\alpha$-enhanced stellar atmosphere models with
\texttt{ATLAS12} code, to extend the present database with
$\alpha$-enhanced stellar spectral libraries. These spectral libraries
will be incorporated in the web interface. 
We will also implement the MARCS atmosphere models \citep{MARCS} into our database.
BCs derived from 3D atmosphere models will also be included in the future.

\section*{Acknowledgements}
We acknowledge the support from the ERC Consolidator Grant funding
scheme ({\em project STARKEY}, G.A. n. 615604). We thank Dr. Angela
Bragaglia and Dr. Diego Bossini for useful discussions on clusters. 
We thank Dr. Bengt Edvardsson for the help on MARCS atmosphere models.
We thank Dr. Xiaodian Chen
for kindly providing us the Python routine computing their extinction
law. This research made use of
Astropy,\footnote{http://www.astropy.org.} a community-developed core
Python package for Astronomy \citep{astropy:2013, astropy:2018}.
We are grateful to the referee for useful suggestions.

\bibliographystyle{aa/aa.bst}
\bibliography{bc} 

\appendix
\section{Usage of the web interface}
\label{app:interface}

The YBC web page stores a user-submitted web form into an input parameter
file, feeds this file as well as the user uploaded data file to the C
executable for obtaining the BCs, and saves the results into an ASCII
file. The BCs will be appended to the end of each line of a
user-uploaded data file. Finally, the web page sends this result file
to the browser for the user to download.

\begin{figure}
  \includegraphics[trim=5 0 5 0,clip,width=0.49\textwidth]{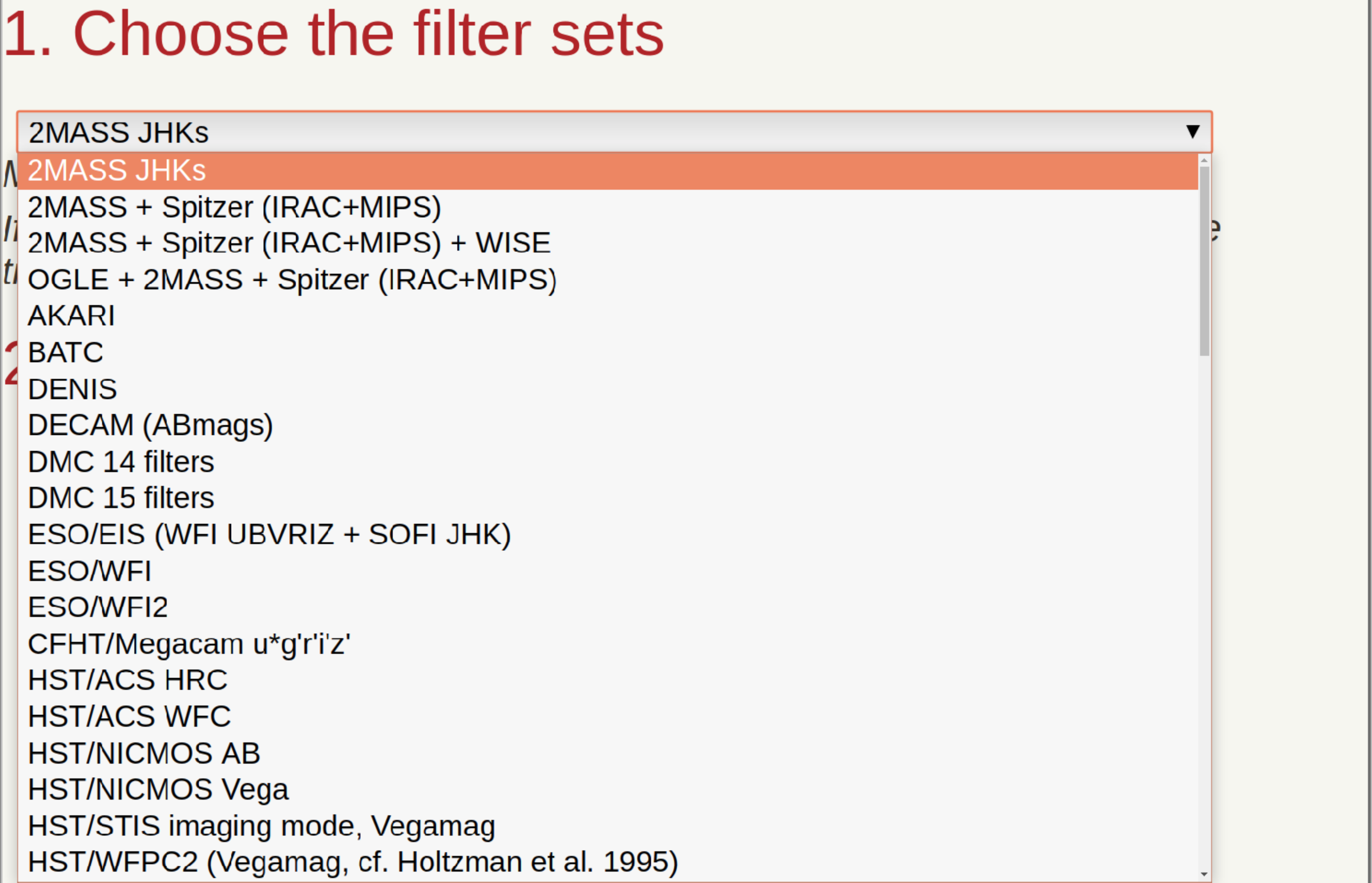}
  \caption{Filter selection step of the web interface. It provides
    more than 70 filter combinations.  It is kept to be updated with
    that of CMD web interface.}
  \label{fig:step1}
\end{figure}

Figure \ref{fig:step1} shows the filter selection step of using the
web interface.  The list of filters come from that of
CMD\footnote{\url{http://stev.oapd.inaf.it/cgi-bin/cmd}.}. It includes
most of the popular filter sets and is frequently expanded.

\begin{figure}
  \includegraphics[trim=8 0 7 0,clip,width=0.49\textwidth]{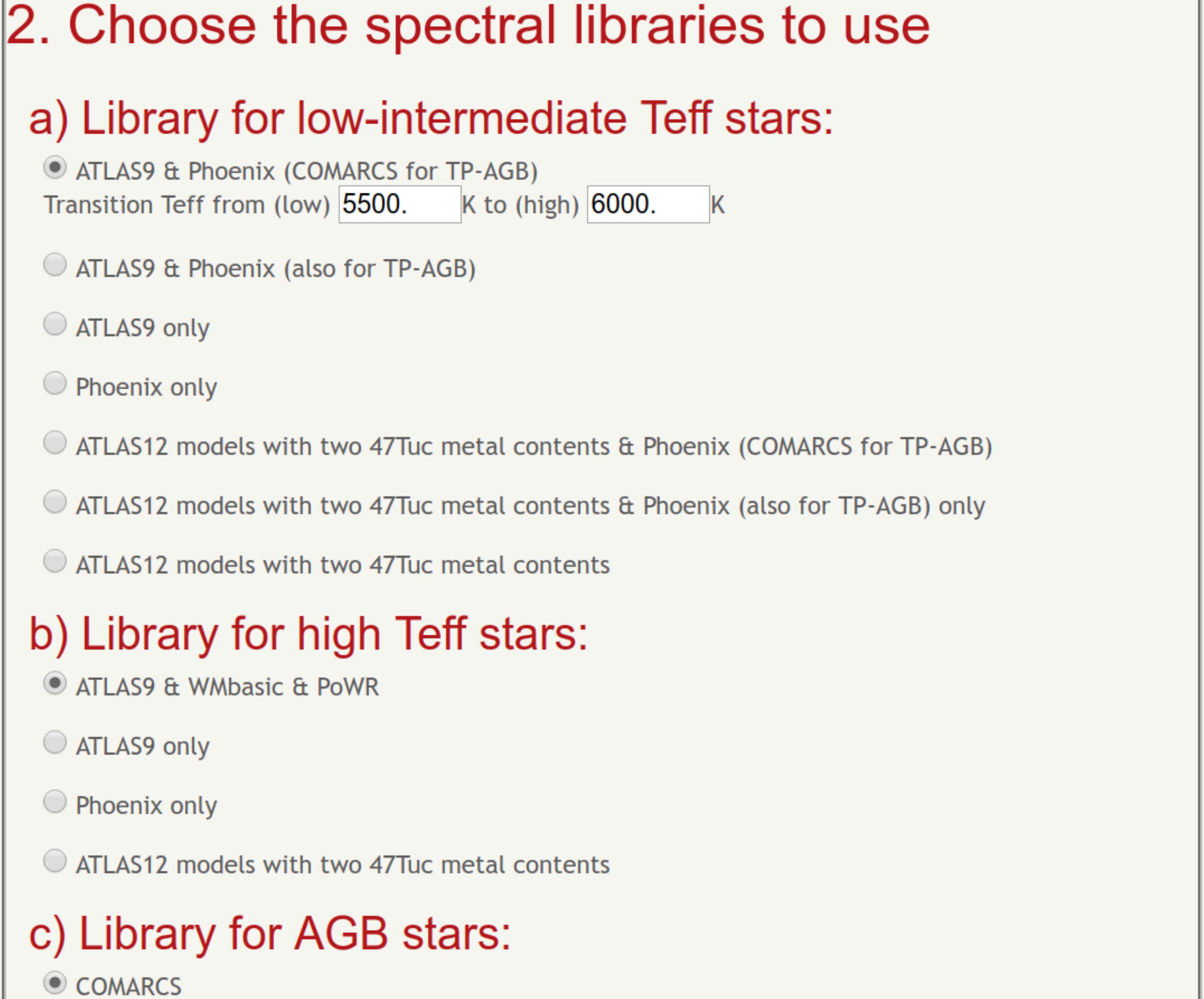}
  \caption{Stellar spectral library selection step. }
  \label{fig:step2}
\end{figure}

Figure \ref{fig:step2} shows the library selection. In section a), the
user has different options for low-intermediate effective temperature
stars. For those options with two libraries, a text element will
show, which specifies the transition $\teff$ between these two
libraries. Section b) is for hot stars and section c) is for AGB
stars.

\begin{figure}
  \includegraphics[trim=0 0 5 0,clip,width=0.49\textwidth]{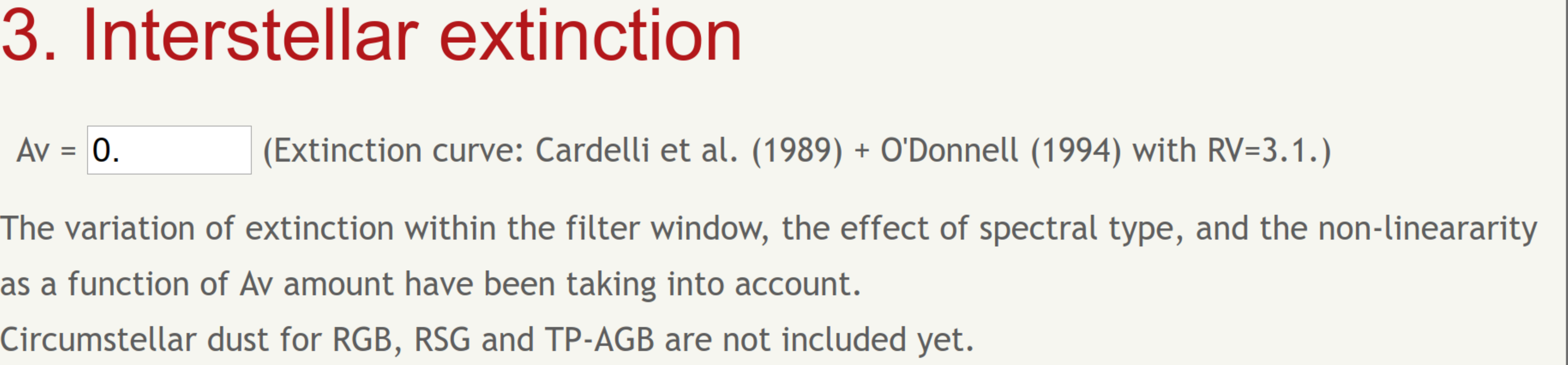}
  \caption{Extinction value setting step. More options will be added in the future.}
  \label{fig:step3}
\end{figure}

In section 3 of the web interface, as shown in Figure~\ref{fig:step3},
the user can specify the interstellar extinction. For the moment, only
the \citet{Cardelli1989} extinction curve with the modification from
\citet{O'Donnell1994} is implemented, but we will soon add more
options. Circumstellar dust for RSG and TP-AGB stars will also be
considered in the next revision.

\begin{figure}
  \includegraphics[width=0.49\textwidth]{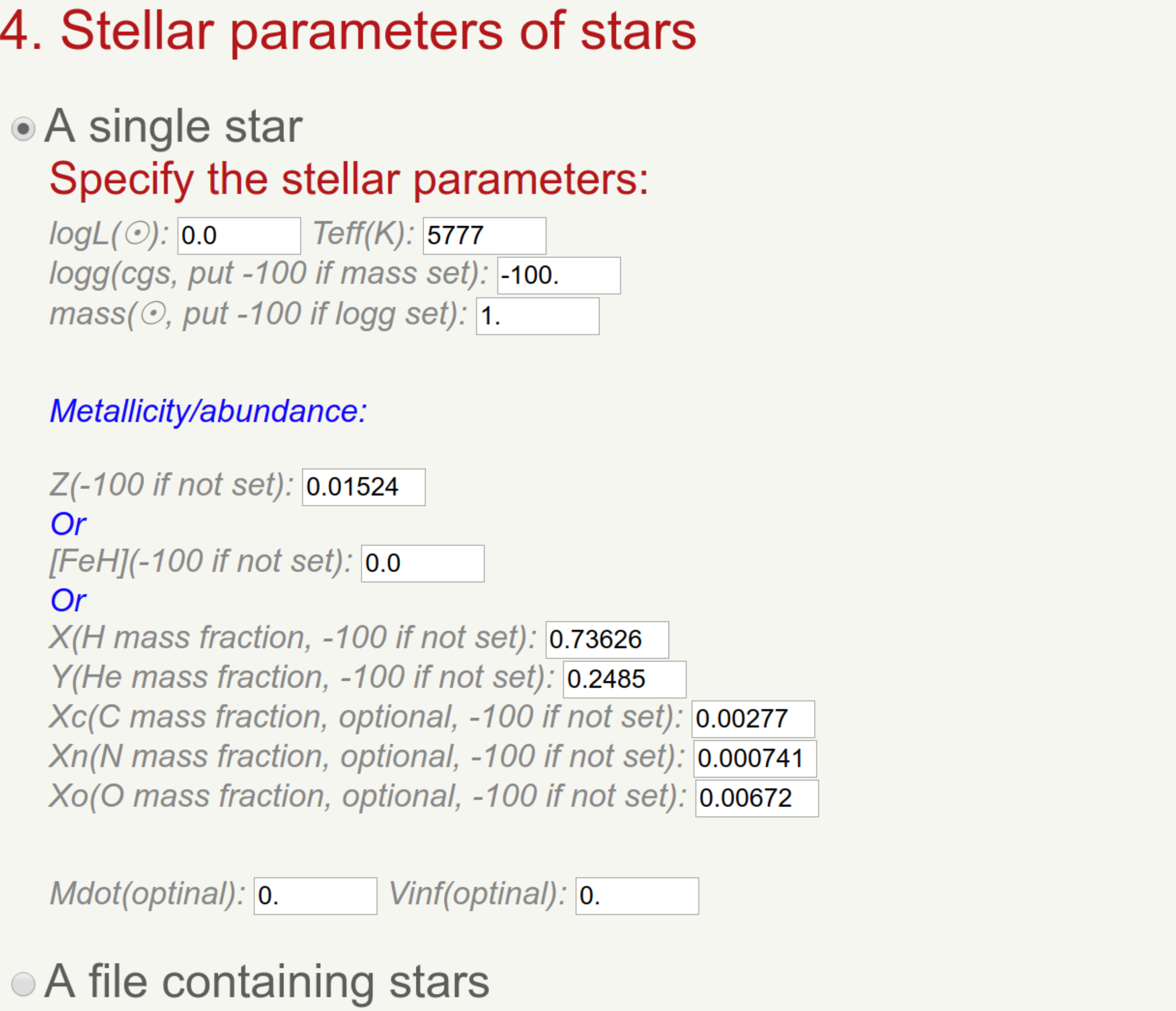}
  \caption{Stellar parameter setting step for a single star.}
  \label{fig:step4}
\end{figure}

\begin{figure}
  \includegraphics[trim=2 0 0 0,clip,width=0.49\textwidth]{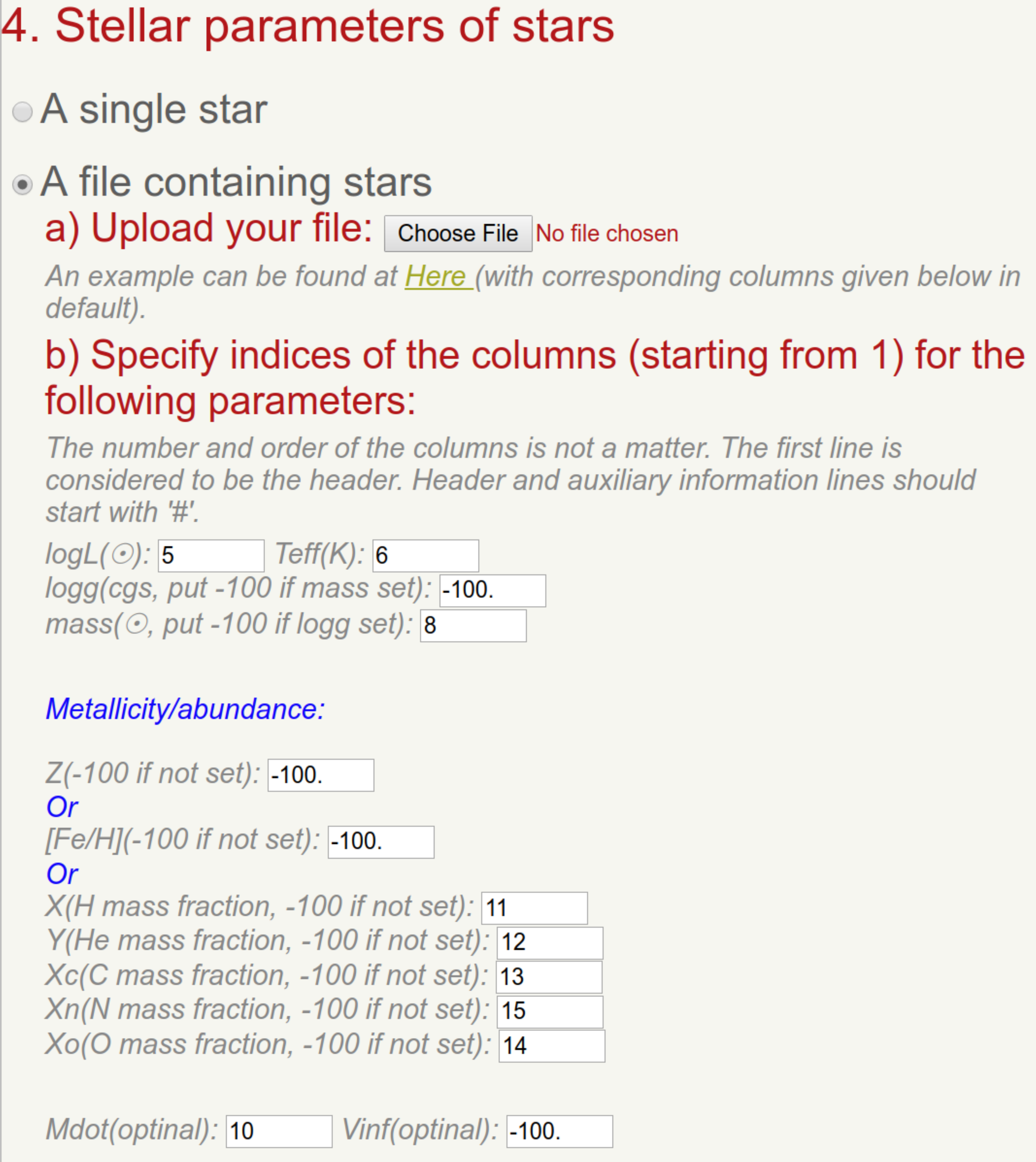}
  \caption{Stellar parameter column specifying for a star catalogue.}
  \label{fig:step5}
\end{figure}

Then, the user can either specify the stellar parameters of a star
(Figure~\ref{fig:step4}) or upload the catalogue
(Figure~\ref{fig:step5}). In the case of uploading the catalogue, the
user has to specify the column number of the stellar parameters. A 1Gby
maximum uploading limit is enforced, and ASCII format is supported.
The surface chemical compositions are used to select the proper stellar spectral
library for the stars.
If the required
chemical abundances are not provided by the user, a solar scaled
abundance is used with the specified metallicity $Z$, which means all
the relevant abundance ratios are the same as those in the Sun. 

Finally, as one clicks on the submit button, a catalogue is generated for download.

\label{lastpage}

\end{document}